\documentclass[preprint2]{proto}
\usepackage{times}
\usepackage{xcolor}
\usepackage{graphicx}
\usepackage{wasysym}
\usepackage{float}
\usepackage{dblfloatfix}

\graphicspath{{./}{Figures/}}

\begin{document}

\title{\textbf{\LARGE Laboratory Experiments to Understand Comets}}

\author {\textbf{\large Olivier Poch}}
\affil{\small\em Université Grenoble Alpes, CNRS, IPAG, France}

\author {\textbf{\large Antoine Pommerol}}
\affil{\small\em Physikalisches Institut, University of Bern, Switzerland}

\author {\textbf{\large Nicolas Fray}}
\affil{\small\em Univ. Paris Est Créteil and Université Paris Cité, CNRS, LISA, F-94010 Créteil, France}

\author {\textbf{\large Bastian Gundlach}}
\affil{\small\em Institut für Geophysik und extraterrestrische Physik, Technische Universität Braunschweig, Germany}

\begin{abstract}

\begin{list}{ } {\rightmargin 1in}
\baselineskip = 11pt
\parindent=1pc
{\small 
In order to understand the origin and evolution of comets, one must decipher the processes that formed and processed cometary ice and dust.  Cometary materials have diverse physical and chemical properties and are mixed in various ways. Laboratory experiments are capable of producing simple to complex analogues of comet-like materials, measuring their properties, and simulating the processes by which their compositions and structures may evolve. The results of laboratory experiments are essential for the interpretations of comet observations and complement theoretical models. They are also necessary for planning future missions to comets. This chapter presents an overview of past and ongoing laboratory experiments exploring how comets were formed and transformed, from the nucleus interior and surface, to the coma. Throughout these sections, the pending questions are highlighted, and the perspectives and prospects for future experiments are discussed.
\\~\\~\\~}%leave this in to get the correct vertical space after the abstract
\end{list}
\end{abstract}  

%-----------------------------------------------------------------------------------------------------------
\section{\textbf{INTRODUCTION}}
\label{sec:intro}

Comets contain clues about the physical and chemical processes that occurred at early times of the solar system formation, and they evolve dramatically when approaching the Sun. They may have brought water, other volatiles and organics to the terrestrial planets. Therefore, fundamental questions of cometary science concern the origin and evolution of comets: What are comets made of? What is their internal structure? How did comets form? How do they evolve, and what causes their activity?

To answer these questions, the cometary science community relies on observations of comets using telescopes (see the chapter by \textit{Bauer et al.} in this volume), flybys or rendezvous space missions (see the chapter by \textit{Snodgrass et al.}), as well as on measurements of cometary samples collected in space or found within Earth atmosphere and surface (see the chapter by \textit{Engrand et al.}). But the proper preparation, analysis and interpretation of these observations also requires theoretical models and experimental analogues, which are based on the current knowledge of comets obtained from previous studies. A strong synergy between observations, theories and experiments is thus essential to understand comets, as well as other objects of the solar system.

The complexity of comets is due to their composition, structure and evolution. Indeed, comets are made of multiple constituents of various origins and volatility (see the chapter by \textit{Bergin et al.}): ices ($\mathrm{H_2O}$, $\mathrm{CO_2}$, $\mathrm{CO}$, $\mathrm{CH_3OH}$, $\mathrm{NH_3}$, $\mathrm{CH_4}$ etc.), semi-volatile (e.g. heavier organic molecules, salts), and refractory materials (refractory organic matter, amorphous and crystalline silicates, carbonates, oxides, sulfides, and metals constituting the dust), the nature and proportion of which may vary depending on the object. Each of these individual constituents can have extremely different physical and chemical properties, and will thus play different roles in the evolution of comets. Of major importance is also the physical arrangement of this mixture of constituents, from the grain scale ($\geq$ 10 nm to 100 µm) to the nucleus scale ($\geq$ 0.1-70 km), to form a low-density agglomerate with a bulk porosity in excess of 50\%. The initial composition and structure of cometary ice and dust depend on comet formation processes and on their place of birth. Both the composition and structure are modified with time, by slow evolutionary processes affecting comets during their long residence in the Oort cloud or the Scattered disk, and during their short active phase when approaching the Sun (see the chapter by \textit{Guilbert-Lepoutre et al.}).

%--------------------------------- Figure comets_vs_expe
\begin{figure*}
\begin{center}
\includegraphics[width=17.1cm]{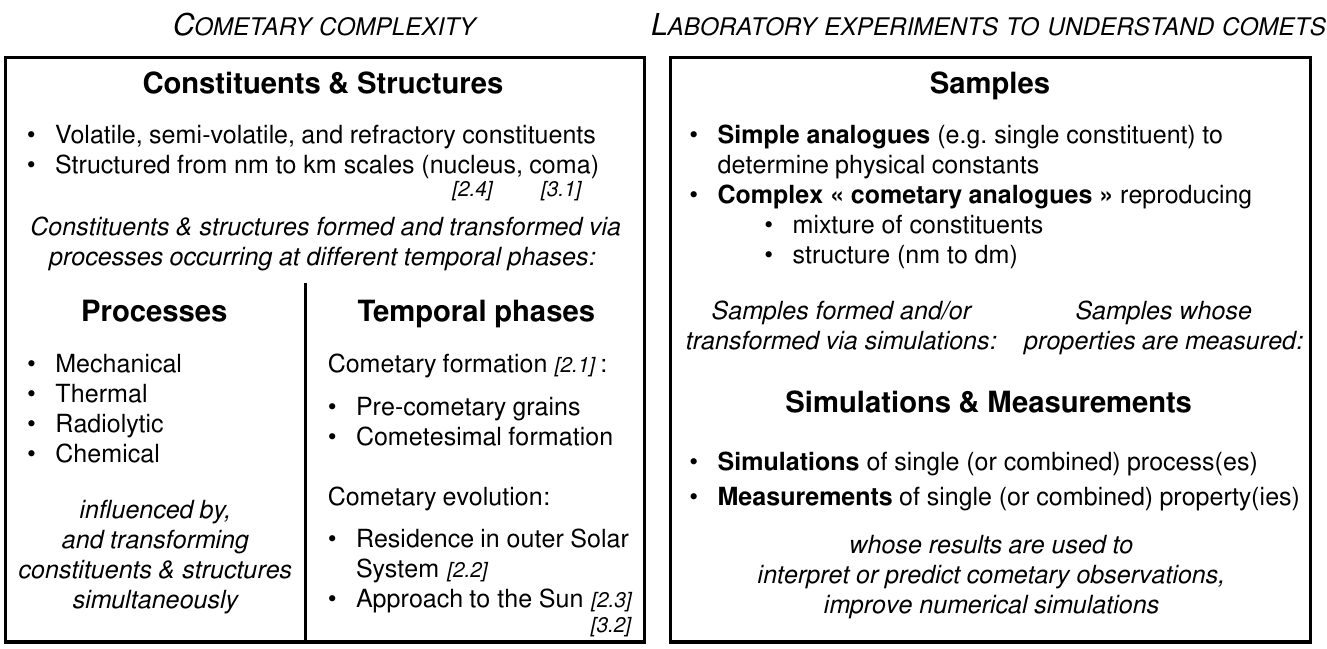}
\caption{This diagram indicates how the natural complexity of comets can be studied and understood using laboratory experiments. Numbers in brackets refer to the sections of this chapter where the relevant experiments are described.}
\label{fig:cometsvsexpe}
\end{center}
\end{figure*}
%---------------------------------

When, where and how did these cometary constituents form? How were they agglomerated together to form cometesimals? What are the fractions inherited from the interstellar medium, transformed, or produced in the proto-planetary disk? How do the composition and structure of comets generate cometary activity? How are they modified by various processes?

To answer these questions, we need laboratory experiments because of the limited ability of theoretical models to simulate the observed natural complexity (Figure \ref{fig:cometsvsexpe}). First principle physics or chemistry arguments cannot predict the evolution of such complex natural systems as cometary materials and structures. In the laboratory, we can produce and study analogues of these cometary constituents and simulate cometary processes.

The objectives of laboratory experiments are to: (1) Improve the general understanding of cometary formation, evolution, and activity, i.e. the physical and chemical processes that governed each of these stages; (2) Interpret observations (light scattering, thermal, dielectric measurements, etc.) of cometary nuclei and comae performed by spacecraft instruments or by telescopes on Earth (or Earth’s orbit), through direct comparison of observations with both theoretical and experimental simulations; (3) Prepare future cometary explorations via predictions of physical and chemical properties.

Two complementary approaches exist for laboratory experiments, depending on the initial complexity of the sample under study (Figure \ref{fig:cometsvsexpe}). Experiments performed with relatively \textit{simple} samples compared to actual cometary material (such as a nm-thick layer of a single constituent, or aggregates of pure silicate grains), are useful to determine physical constants and their dependencies on sample properties that can be implemented in theoretical models. On the other hand, experiments performed on more \textit{complex} samples, often called “\textit{cometary analogues}” are used to simulate some aspects of cometary complexity (mixture of constituents, structure, combination of processes etc.) and assess their influence on the processes (thermal, radiolytic, mechanical processes etc.) modifying comets. The measurements done on these complex cometary analogues can be directly compared to observations, and/or can be used to test theoretical models. In all cases, it is crucial to characterize the composition and structure of the produced samples in quantitative detail before they are used for experiments.

This chapter presents an overview of laboratory experiments relevant to comets. Section 2 describes experiments used to understand cometary nuclei properties, from their formation to evolution during residence in the outer solar system and approach to the Sun. Section 3 covers experiments exploring the properties and evolution of comae particles, and section 4 provides some perspectives on experiments needed in the future.

%-----------------------------------------------------------------------------------------------------------
\section{\textbf{COMETARY NUCLEI}}
\label{sec:nextsection}

%---------------------------------
\subsection{Experiments on comet formation, from interstellar and circumstellar grains to cometesimals}
\label{sec:subsection10}

\subsubsection{Brief summary on comet origin and evolution}

Comets formed in the young solar system when the protoplanetary disk was still present (see the review by \citet{weissman2020a}, and the chapter by \textit{Aikawa et al.} in this volume). However, it is not yet known precisely where and when comets have formed. In this section 2.1, we refer to “\textit{cometesimals}” as the building blocks that formed comets. After their formation, comets have been altered by cometary evolutionary processes (see Figure \ref{fig:cometsvsexpe}, and section 2.2 and onward).

Due to the presence of water ice and more volatile species, called super-volatiles (i.e., having an equilibrium sublimation pressure higher than $10^{-9}$ bar at 100~K, like $\mathrm{CO_2}$, $\mathrm{NO}$ etc.) and hyper-volatiles (i.e., higher than $10^{-9}$ bar at 50~K, like $\mathrm{Kr}$, $\mathrm{CO}$ etc.) \citep{fray2009a}, we know that cometesimal formation must have taken place beyond the snow line, which describes at which distance from the central star water ice (or other frozen volatile species) are stable in the protoplanetary disk \citep{stevenson1988}. The average low density of cometary nuclei (below 1 g cm${}^-{}^3$ whenever estimated), favours two possible scenarios for their formation, namely the gravitational instability scenario that predicts a cometary material that consists of pebbles and fractal agglomerates, and the mass transfer scenario that predicts an homogeneous material that consists of (sub-)micrometer-sized grains (see the chapter by \textit{Simon et al.} in this volume). Laboratory experiments are performed with the aim to study these growth processes via coagulation of grains and collision of aggregates having different compositions. The formation scenario may also influence the composition of cometesimals, because the direct growth scenario requires more time than the gravitational instability one. Indeed, if cometesimals formed relatively fast after the first solid components of the solar system (the Calcium–Aluminum-rich Inclusions, CAIs, found in chondrite meteorites), they must have experienced significant heating due to the decay of short-lived radionuclides \citep[see, e.g.,][]{Prialniketal2008}, leading to thermally-induced chemical reactions and depletion of the volatiles in the interior and probably in the whole object. In contrast, a delayed formation of cometesimals would imply a lower thermal evolution induced by radioactive decay than an early formation.

Observations and analyses of cometary materials indicate that grains of ice and dust of very different origins were mixed to form cometesimals \citep{brownlee2014a}. Some of these grains might have formed in molecular clouds or dense cores (pre-stellar environments, also called “\textit{interstellar medium}” abbreviated ISM) and incorporated intact in cometesimals (see the chapter by \textit{Bergin et al.} in this volume), and many others have been modified, or (re-)formed in various regions of the protostellar and protoplanetary disk before being transported and incorporated in cometesimals (see the chapters by \textit{Aikawa et al.} and \textit{Engrand et al.} in this volume). Moreover, possible later accretion of refractory materials after the icy cometesimal formed cannot be excluded (see the chapter by \textit{Engrand et al.}). The paragraphs below present the different pathways to synthesize and study analogues of cometary volatile molecules, carbonaceous and organic dust, as well as minerals via laboratory experiments.

Important open questions about the chemical composition of cometesimals are: the degree of inheritance versus reset of the volatile and refractory materials from pre-stellar and proto-stellar phases to cometesimals, and the elemental budgets throughout this evolution, i.e. the identification of the materials/molecules carrying each atom C, O, N and S to account for their cosmic abundance (see reviews in \citet{caselli2012a, boogert2015a, oeberg2021a, boogert_questions_2019}). Laboratory experiments, in conjunction with theoretical works, play important roles in answering these questions, by providing means to identify and quantify the volatile and refractory species from observational data of pre-cometary environments, and by studying the processes affecting them before their incorporation in cometesimals.

%--------------------------------- Figure IceChemistry
\begin{figure*}
\begin{center}
\includegraphics[width=17.1cm]{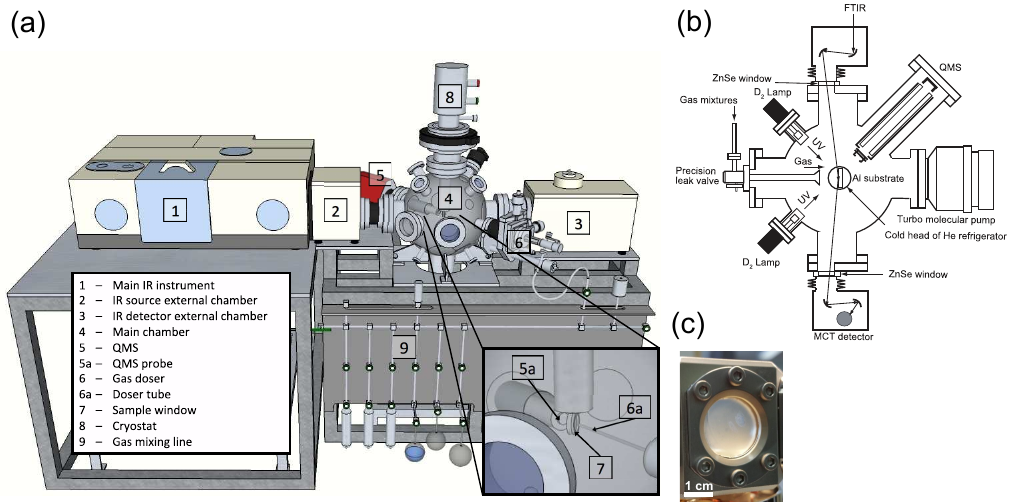}
\caption{Setups for astrochemistry of (pre-)cometary ices analogues, deposited as thin (nm-$\mu$m) layers. (a) Ultra-high vacuum and cryogenic setup, with sample monitoring by infrared spectroscopy and mass spectrometry \citep{lauck2015}. (b) Another setup, with UV irradiation of the sample \citep{oba2017a}. Various irradiation and monitoring techniques can be used. (c) Refractory organic residue on a sample substrate \citep{aug2016a}. Reproduced by permission of the AAS.}
\label{fig:icechem}
\end{center}
\end{figure*}
%---------------------------------

\subsubsection{Chemistry of pre-cometary grains}

\subsubsubsection{2.1.2.1. Volatile molecules}

Volatile molecules found in comets are mainly $\mathrm{H_2O}$, $\mathrm{CO}$, $\mathrm{CO_2}$, and other minor constituents such as $\mathrm{CH_3OH}$, $\mathrm{NH_3}$, $\mathrm{CH_4}$, etc. including larger carbon-bearing molecules which are called “\textit{organic molecules}” or “\textit{complex organic molecules}” (COM), the latter denomination being used in ISM studies to refer to all molecules containing at least 6 atoms including carbon (see \citet{herbst2009a} and the chapter by \textit{Bergin et al.} in this volume). These molecules are formed by chemical reactions in the gas phase, at the surfaces of bare dust particles, or in the ice mantles that cover these particles in cold dense interstellar clouds and possibly in later stages of star formation (see reviews in \citet{herbst2009a, herbst2014a}). To predict or interpret observations and to complete numerical models, laboratory experiments are performed to obtain parameters (rate constants, branching ratios) of these chemical reactions in the gas \citep{smith2011a}, gas-solid, and solid phases \citep{oeberg2009a}. Other experiments also provide means to identify and quantify these molecules, via gas phase molecular excitation, and spectroscopy of gases and ices at infrared wavelengths (see references in \citet{herbst2009a, herbst2014a, oeberg2021a}). These experiments dedicated to the chemistry of volatile molecules in pre- and protostellar environments are crucial to understand their inheritance and/or processing between the ISM and comets (see reviews in \citet{caselli2012a, boogert2015a, oeberg2021a}, and the chapters by \textit{Bergin et al.} and \textit{Aikawa et al.} in this volume).
	
Of particular interest to understand the composition of cometesimals are the experiments investigating the evolution of ice mantles coating interstellar grains from quiescent clouds, to young stellar objects envelopes and disks \citep{caselli2012a}. Chemical reactions in and on these ices take place through neutral-neutral atom and radical addition reactions following the accretion of species, or via the interaction with ultraviolet photons and impacting particles (protons, ions, electrons from galactic cosmic rays), and also via heating. Laboratory studies aiming at reproducing the astrophysical conditions experienced by ice mantles are described below (more detailed reviews can be found in \citet{linnartz2015a, oeberg2016a, gudipati2015a}).

\textit{Setups for ice mantle chemistry.} Figure \ref{fig:icechem} shows setups consisting of an ultra-high vacuum chamber (with pressure down to 10$^{-10}$ mbar) containing a metal or infrared-transparent salt surface ($\sim$2 cm diameter) mounted on the tip of a Helium cryostat in order to be cooled down to about 10 K, and controlled by resistance heaters. This cold substrate is exposed to a flow of gas of controlled composition either pure or a mixture ($\mathrm{H_2O}$, $\mathrm{NH_3}$ etc.), to build a layer of ice few nanometers to micrometers thick. This ice surface is then exposed to heat, by changing the temperature of the cryostat, and/or to vacuum ultraviolet photons (from 115 to 200 nm, using a microwave driven $\mathrm{H_2}$ discharge lamp), X-rays (from a synchrotron), energetic particles (neutral or ionised atoms, electrons, e.g. from an accelerator) inducing desorption, structural changes and chemical reactions. These changes are monitored \textit{in situ}, most often via infrared spectroscopy (transmission or reflectance Fourier transform infrared, FTIR) and mass spectrometry (quadrupole or time of flight mass spectrometer, QMS or TOFMS) during the irradiation and/or during progressive heating of the substrate (called temperature programmed desorption, TPD). Infrared spectroscopy monitors the composition of the solid phase by identifying absorption bands at frequencies specific to molecular vibrations in the mid-infrared (MIR, 2.5–25 $\mu$m). In some experiments, \textit{in situ} ultraviolet-visible (UV-Vis, 190-700 nm), vacuum ultraviolet (VUV, 100-200 nm), Raman, far-infrared and THz time-domain spectroscopies are also performed \citep{allodi2013a}. Mass spectrometry techniques use an ionization source, a mass-selective analyzer, and an ion detector to identify individual molecules produced in the gas phase and/or in the ice. During progressive TPD heating (rate from 0.1 to 10 K min$^{-1}$), each molecule has its maximum sublimation rate at a specific temperature, facilitating its identification by mass spectrometry \citep{oeberg2009a}. The temperature at which the molecule sublimates also depends on its mixing, trapping, or segregation, providing information not only on the composition but also on the structure of the solid phase. Some experiments implement \textit{in situ} analysis of the ice via non-thermal laser desorption/ablation and ionization mass spectrometry (LD-TOFMS or Matrix Assisted Laser Desorption Ionization, MALDI-TOFMS, \citet{henderson2015a}). 

\begin{figure*}
\begin{center}
\includegraphics[width=17.1cm]{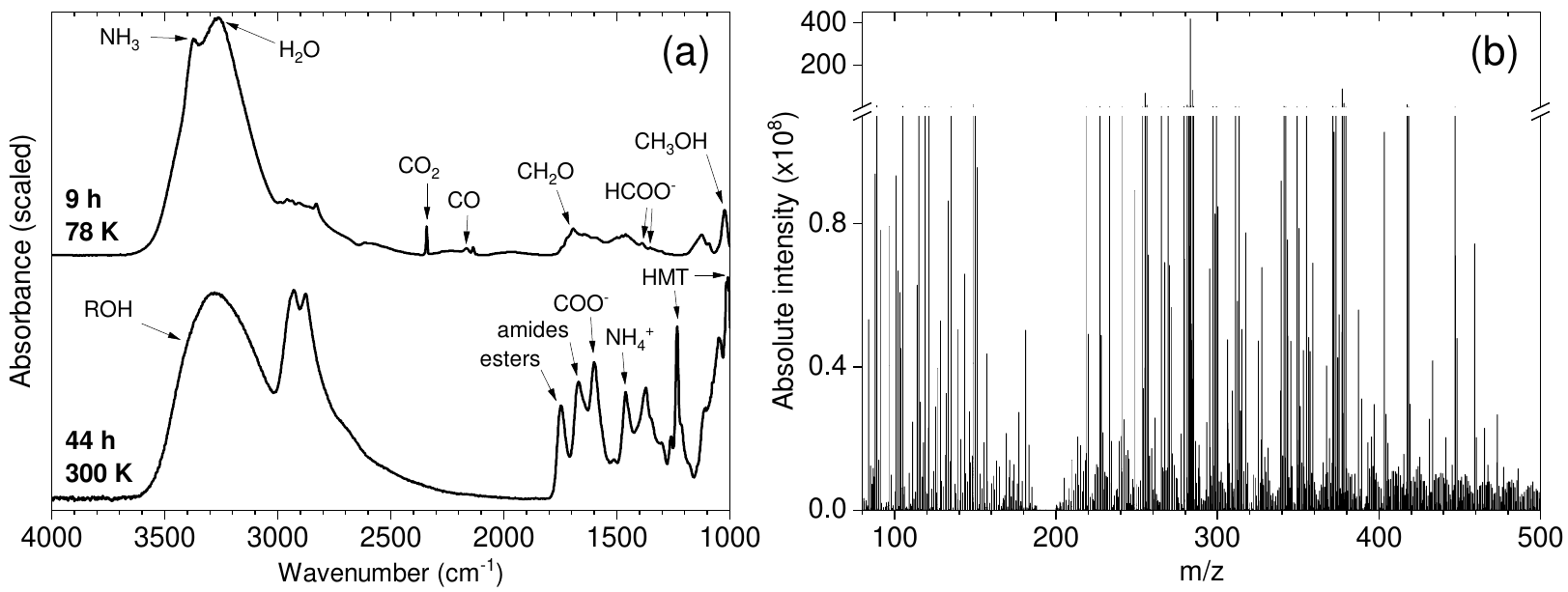}
\caption{Molecular diversity after irradiation of ices and warming. (a) FTIR spectra of (top) $\mathrm{H_2O}$:$\mathrm{NH_3}$:$\mathrm{CH_3OH}$ in a ratio 3:1:1 after about 9 h of UV irradiation at 78 K, (bottom) the resulting organic residue after 44 h of UV at 78 K and warmed up to 300 K. (b) Mass spectrum of the methanol-soluble fraction of this organic residue, showing thousands of peaks. \citep{danger2013a}}
\label{fig:danger13}
\end{center}
\end{figure*}

\textit{Results of ice mantle chemistry experiments.} Such experiments have shown that the exposure of ices to ionizing radiation (UV, energetic particles) breaks molecular bonds and creates radicals and ions, which can diffuse, meet, and react in or on the ice to form more complex molecules. Photolysis or radiolysis induces similar products, suggesting that the overall chemical evolution mainly depends on the amount –not the kind– of energy delivered to the ice mantle \citep{gerakines2001, gerakines2004, rothard2017a}. Heating in star-forming regions (from 10 to 100 K or more) facilitates the mobility of these species produced by photolysis or radiolysis, allowing their reaction. Moreover, mixtures of neutral molecules ($\mathrm{NH_3}$, $\mathrm{CO_2}$, etc.) can produce new molecules by purely thermal reactions (acid-base, nucleophilic addition and elimination, or condensation), without any other source of energy, providing that the initial abundance of the reactants, the temperature, and the time available are adequate (see review in \citet{theul2013a}). The association of ionizing radiations followed by warming of a mixture of ices results in an impressive molecular diversity (Figure \ref{fig:danger13}). Water, which is the most abundant molecule in pre-cometary grains and comets, plays important roles in this chemistry, as reactant, as catalyst, or as trapping matrix and diffusion surface allowing reactions to occur (\citet{fresneau_trapping_2014} and references therein, \citet{ghesqui2018a}). Newly formed molecules, of larger size, are often less volatile (more refractory) than their precursors. After warm-up and complete sublimation of the ice at the end of the experiment, the molecules remaining in solid phase at room temperature (Figure \ref{fig:icechem}c) are thus called the “\textit{refractory residue}” (historically called “\textit{yellow stuff}” in \citet{greenberg1995a} and references therein). These are semi-volatile molecules, such as salts (ionic solids), including molecules of prebiotic interest, as well as polymers up to refractory carbon-bearing compounds. \textit{Ex-situ} analyses, outside the simulation chamber, are performed via various techniques to reveal the nature of this residue. Experiments on the chemistry of refractory residues are presented in the next section (2.1.2.2).

\textit{Diffusion and reaction parameters.} Laboratory experiments with single-component or binary ices enable the determination of reaction pathways and physical constants, such as reaction rates, cross-sections, desorption rates and branching ratios needed to complete existing chemical networks used to compute the evolution of the ice/gas composition (for a review of these models, see section 1.3.2 in \citet{oeberg2021a}). These parameters are significantly influenced by the structure of the ice (porosity, morphology etc.), so experiments on various ice structures (produced by varying the deposition method on the cold substrate, or the temperature) should be performed \citep{isokoski2014, noble2020, kouchi2021}. Moreover, the simultaneous presence of several molecules in the ice also changes the reaction kinetics compared to pure or simple ice mixtures. Other important parameters to obtain via experiments are diffusion rates of different radicals, ions or neutrals in ices at different temperatures \citep{oeberg2016a, mispelaer2013a, minissale2019a} (see next section 2.3.4.2).

\textit{Influence of dust on the ice chemistry.} Pre-cometary ices have condensed on refractory particles, which are made of minerals (mainly silicates, etc.) and/or carbonaceous materials. The nature of the refractory particle (mineral/carbonaceous) and its structure (porosity, etc.) will affect the electrostatic or bonding interactions with the ice molecules, the energy dissipation of exothermic reactions, the active area in contact with the molecules, influencing reaction rates, diffusion and concentration of species. Only a limited number of studies have addressed the effects of dust substrates on the astrophysical ice chemistry, but some studies demonstrated the catalytic effect of dust surface on processes in ices (see for example \citet{potapov2019a}, and review in \citet{jaeger2019a}). In the setup described above (Figure \ref{fig:icechem}), ices are deposited as a layer on a substrate. However, observations suggest that cometary dust particles possess a large range of porosity, from highly porous to compact aggregates \citep{wooden2008a, levasseur-regourd2018a}. Ices condensing on such aggregates probably fill the pores, having a much larger surface of contact with the dust than in typical experiments, and consequently enhanced interactions. Future laboratory experiments should thus develop methods to prepare such ice/dust mixtures, with various composition and structure, and analyze their influences on ice chemistry and spectroscopy.

%--------------------------------- Figure Danger13
%---------------------------------

\subsubsubsection{2.1.2.2. From ices to refractory organic dust} The organic refractory residues, formed after energetic irradiation followed by warming and sublimation of ices (Figure \ref{fig:icechem}c), comprise an impressive diversity of molecules, such as salts (ionic solids), individual molecules, polymers and other macromolecules, including some of prebiotic interest, which are analyzed by different techniques available in the laboratory.

\textit{Analysis of the entire molecular diversity.} After energetic processing, warming and subsequent sublimation of the ice sample inside the experimental setup, the refractory residue remaining on the substrate can be analyzed \textit{in situ}, most often via infrared spectroscopy, and/or \textit{ex situ} via mass spectrometry following desorption or extraction techniques (Figure \ref{fig:danger13}). Other \textit{ex situ} analyses such as X-ray absorption near-edge structure spectroscopy (XANES), energy-dispersive X-ray spectroscopy (EDX), and transmission and scanning electron microscopy (TEM and SEM, respectively) are also sometimes used for the elemental (N/C, O/C etc.), chemical (bonds, chemical functions) and structural (macromolecular organization) characterization of residues.
Infrared spectroscopy of residues has revealed the presence of several molecular species and chemical groups in the residues, in particular hexamethylenetetramine (HMT, $\mathrm{[(CH_2)_6N_4]}$), ammonium salts of carboxylic acids $\mathrm{[R}$–$\mathrm{COO^-}$ $\mathrm{NH_4^+]}$, amides $\mathrm{[H_2N}$-$\mathrm{C(}$=$\mathrm{O)}$-$\mathrm{R]}$, esters $\mathrm{[R}$-$\mathrm{C(}$=$\mathrm{O)}$-$\mathrm{O}$-$\mathrm{R]}$ and species related to polyoxymethylene (POM, $\mathrm{[(-CH_2O-)_n]}$) \citep{munoz2003a}. HMT, a nitrogen-bearing cyclic molecule, is an abundant component of some refractory residues obtained after processing of various initial ice mixtures: for example, after irradiation (via UV or protons) and warming of $\mathrm{H_2O}$:$\mathrm{NH_3}$:$\mathrm{CH_3OH}$ ices \citep{bernstein1995a, cottin2001a, munoz2003a, munoz2004a, oba2017a} (Figure \ref{fig:danger13}a), or after only warming of $\mathrm{H_2CO}$:$\mathrm{NH_3}$:$\mathrm{HCOOH}$ or $\mathrm{CH_2NH}$:$\mathrm{HCOOH}$ ices \citep{vinogradoff2013a}. HMT can reach up to 50 wt\% of the total organic products, especially if methanol ($\mathrm{CH_3OH}$) is present in the initial ice mixture \citep{danger2013a}. Depending on the relative concentration of $\mathrm{NH_3}$ or $\mathrm{CN^-}$ and formaldehyde ($\mathrm{H_2CO}$) in the ice mixture, $\mathrm{H_2CO}$ can polymerize to form polyoxymethylene (POM) or POM-like polymers which are then the most abundant components of the residues \citep{butscher2019a, duvernay2014a}. These experiments have suggested the presence of HMT or POM in comets. Although HMT is found at part-per-billion level concentration in carbonaceous chondrites \citep{oba2020}, it has not been detected in a comet up to now, and the presence of POM in Giotto \citep{huebner1987a} and Rosetta \citep{wright2015a} observations is debated \citep{altwegg2017a}. These compounds, although abundant in some laboratory refractory residues, are not observed in comets yet, either because they are mainly decomposed, or because their precursors are not formed or destroyed, under cometary conditions. Future experiments should investigate these questions and suggest when/where these compounds could form and be preserved, or under which conditions they decompose. 
Other abundant components of refractory residues primarily characterized by their infrared absorption spectra are ammonium salts ($\mathrm{NH_4^+ X^-}$, where $\mathrm{X^-}$ is the anion of a generic acid molecule $\mathrm{XH}$). They can be produced by acid-base or nucleophilic addition reactions activated by heat \citep{theul2013a}. If $\mathrm{NH_3}$ and an acid molecule (for example $\mathrm{HCOOH}$, $\mathrm{HCN}$, $\mathrm{HNCO}$ etc.) can diffuse close enough to each other to exchange a proton, the pair of ions $\mathrm{NH_4^+}$ (ammonium) and $\mathrm{HCOO^-}$ (formate) is formed. Alternatively, additions of $\mathrm{NH_3}$ on $\mathrm{CO_2}$ can end-up in the formation of ammonium carbamates \citep{theul2013a}. Laboratory experiments performed for many years had predicted the presence of ammonium salts as a likely component of comets \citep{colangeli2004}, which was confirmed in the dust of comet 67P by the VIRTIS and ROSINA instruments of Rosetta \citep{altwegg2020a, poch2020}. Moreover, $\mathrm{NH_4^+}$ is suspected to be present in interstellar ices, as a counter-ion of isocyanide $\mathrm{OCN^-}$ which has been identified in interstellar ices \citep{oberg2011, boogert2015a, boogert2022, mcclure2023} and in different protostar environments \citep{schutte2003a, vanbroekhuizen2005}. Therefore, ammonium salts might be inherited from interstellar ices, and/or they may form at later cometary stages (see section 2.3.4.2).

For \textit{ex situ} analyses, the organic refractory residue is collected and placed in the chamber of a microprobe laser-desorption laser-ionization mass spectrometer \citep{dworkin2004a} or a laser-desorption time-of-flight mass spectrometer (LD-TOFMS) \citep{modica2012a}. LD-TOFMS analyses have revealed the presence of macromolecular components between 1000 and 3000 amu in residues \citep{modica2012a}. Other mass spectrometers include (Very) High Resolution Mass Spectrometry (HRMS or VHRMS) techniques, such as Fourier Transform Ion Cyclotron Resonance (FT-ICR) or Fourier Transform Orbitrap (FT-Orbitrap). For these analyses, the refractory residues are dissolved (often in ultrapure methanol) before being injected into the mass spectrometer (Figure \ref{fig:danger13}). VHRMS provides sufficient mass resolution to retrieve the chemical formulae of the molecules, enabling the sorting of molecules in families according to their content in specific atoms, group of atoms, and even aromaticity \citep{danger2013a, danger2016a, fresneau2017a, gautier2020a}. VHRMS is a promising technique to determine how the overall chemistry of analogues is affected by different production conditions, and how it compares to that of cosmo-material organic matter, especially future comet samples that will be returned on Earth (see the chapter by \textit{Snodgrass et al.} in this volume).

\textit{Identification of specific molecules.} To search for specific molecules in refractory organic residues, chromatography techniques are used in order to separate the molecules according to their chemical properties, before they are weighed by mass spectrometry. After dissolution, residues are analyzed by high performance liquid chromatrography coupled to a mass spectrometer (HPLC-MS). Residues can also be analyzed via gas chromatography coupled to mass spectrometry (GCMS), providing that they are treated chemically to increase the volatility of some molecules (a step called derivatization or functionalization). These techniques allowed for the confirmation of the presence of HMT \citep{cottin2001a, munoz2004a}, amines, and various nitrogen- and oxygen-bearing linear or cyclic organic compounds including precursors of molecules of prebiotic interest such as urea, hydantoin and carbamic acid \citep{chen2007a, marcellus2011a, nuevo2010a}. Molecules of prebiotic interest identified in refractory residues are amino acids \citep{bernstein2002a, nuevo2008a}, amphiphiles \citep{dworkin2001a}, nucleobases \citep{oba2019a}, and sugars \citep{meinert2016a, nuevo2018a}. Moreover, studies using enantioselective multidimensional gas chromatography coupled to time-of-flight mass spectrometry (GC×GC-TOFMS) have shown that the circular polarization of the VUV light used to irradiate the ices induces the production of a small enantiomeric excess in amino acids \citep{marcellus2011a, modica2014a}. This result suggests that an enantiomeric excess brought to the early Earth by cometary materials might have been amplified via prebiotic chemistry, potentially explaining the specific amino acids enantiomers chosen by terrestrial life forms whose homochirality is a universal property. Among the molecules of prebiotic interest, only glycine and precursors such as ammonium salts have been firmly identified on comets \citep{altwegg2016a, poch2020}, but neither of these molecules are chiral. These techniques employed on analogues will allow searching for these and other molecules in future samples returned from comets, and some of them may be miniaturized to be run onboard future space missions. The main drawback of \textit{ex situ} analyses is the exposure of the organic residue to air and ambient temperature, and the possible alteration of the most reactive species it contains with water vapor or $\mathrm{O_2}$ for example. Future laboratory experiments should address this problem, by developing new methods to transfer the samples for \textit{ex situ} analyses, or by developing novel techniques for \textit{in situ} analyses, maybe inspired by other research fields \citep{fulvio2021a}. 

\textit{From refractory ice residues to even more refractory compounds.} The carbon in comet 67P, and possibly in other comets, appears to be mainly present as high-molecular-weight organic matter, analogous to the insoluble organic matter (IOM) found in the carbonaceous chondrite meteorites \citep{fray2016a, bardyn2017}. The refractory organic matter of comet 67P has a higher hydrogen/carbon ratio than IOM, suggesting that it could be more pristine \citep{isnard2019}. How this refractory organic matter was formed is a major question of cosmochemistry. Organic refractory residues obtained after sublimation of the ices may have been exposed to irradiation in pre- and proto- stellar environments, as well as heat in the forming protoplanetary disk. These subsequent processings of organic residues are studied by dedicated laboratory experiments, described in this paragraph. Moreover, irradiations and heat could also transform some of the organic compounds after their incorporation in cometesimals, as discussed in sections 2.2.3 and 2.3.4.2. Experiments have shown that partial thermal processing at 300-400°C is required to convert organic residues made from ice photoprocessing and warm-up into the amorphous carbon with low heteroatom content found in IDPs \citep{munoz2006a}. Organic residues subjected to further UV irradiation produced a thin altered dark crust insoluble, on top of the initially soluble residue \citep{demarcellus2017a, piani2017a}. This insoluble material shows spectral similarities with natural samples of IOM extracted from carbonaceous chondrites \citep{demarcellus2017a}. TEM and SEM morphological observations showed amorphous nanospherules similar to organic nanoglobules observed in the least-altered chondrites, chondritic porous interplanetary dust particles (CP-IDPs), and cometary samples \citep{piani2017a}. However, to date no experiment has been able to reproduce all the structural, chemical and isotopic properties of IOM. \citet{faure2021} performed ion irradiation experiments of various organic polymers, revealing a dramatic precursor effect on the final chemistry of the residues below a nuclear dose of $\sim$10 eV.atom$^{‐1}$. Above this dose, any precursor transforms into amorphous carbon. Other experiments exposed refractory organics residues, previously prepared in the laboratory, to solar UV from low-Earth orbit \citep{baratta2019a, greenberg1995a}. Future exposure experiments in outer space (e.g. on the planned “\textit{Lunar Gateway}” space station) may allow simultaneous exposure of many different samples to the solar radiations on long time scales \citep{cottin2017a}. However, these experiments on thin layers only produce very low amounts of organic residues, preventing their use in complex cometary analogues for larger scale experiments (see sections 2.3.2, 2.4). For such experiments, an alternative is to produce several hundreds of mg of analogues of these carbonized organic residues by thermal degradation of \textit{tholins} produced from gases, or HCN polymers produced from aqueous chemistry \citep{bonnet2015}.

%--------------------------------- Figure Nuth2016 - OrganicGas
\begin{figure*}[ht]
\begin{center}
\includegraphics[width=17.1cm]{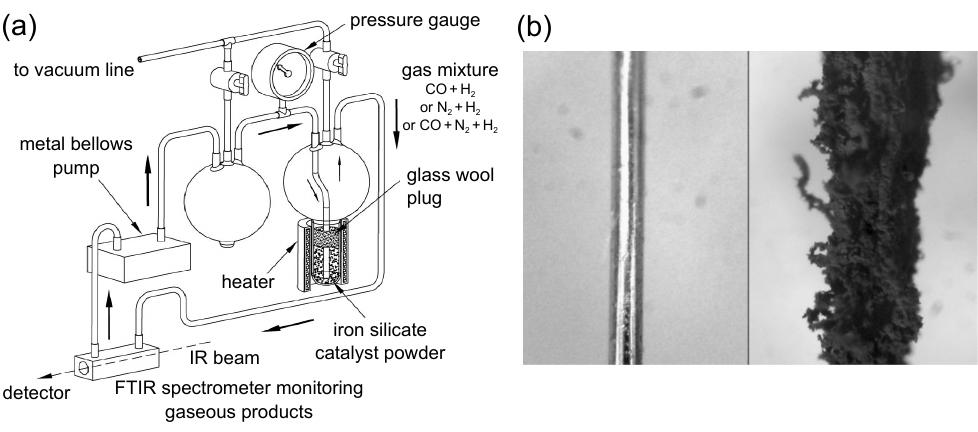}
\caption{Production of organic dust via catalytic reduction of $\mathrm{N_2}$ and $\mathrm{CO}$ by $\mathrm{H_2}$ on mineral grains or on an iron wire ($\sim$200 $\mu$m in diameter) \citep{nuth2008a, nuth2016}. Such dust is used as a cometary analogue in light scattering experiments (see Fig. \ref{fig:lightscat}). Figures reproduced from \citet{nuth2020} with permission.}
\label{fig:organicgas}
\end{center}
\end{figure*}
%---------------------------------

\subsubsubsection{2.1.2.3. Amorphous carbon grains}
In the diffuse ISM, a significant portion of carbon is expected to be in the dust \citep{dishoeck2014}. Infrared spectra of the diffuse interstellar medium suggest that this carbon dust is much more similar to carbon-rich hydrocarbon materials than energetically processed ice residues containing heteroatoms \citep{pendleton2002a}. \citet{henning2004a} and \citet{ehrenfreund2010a} provide reviews of laboratory studies on the formation, evolution and means of characterization of different forms of carbon-rich dust expected in the ISM. This carbon dust could be in the form of various cross-linked three-dimensional networks of “\textit{amorphous carbon}” that contain only C and H, and “\textit{hydrogenated amorphous carbon}” containing higher level of H. Molecules such as polycyclic aromatic hydrocarbons (PAHs) and their clusters, widespread in the ISM, could contribute to the formation of this carbon dust and/or be produced from their decomposition. Finally, “\textit{macromolecular organic matter}” containing additional atoms such as O, N and S, may be produced through processing of these carbon dust precursors, or from ice chemistry as discussed above (section 2.1.2.2). It is unclear if this interstellar carbonaceous dust was preserved, modified, or destroyed in the protoplanetary disk, and how much could have been incorporated in cometesimals.

Laboratory experiments have been carried out to study amorphous carbon with different degrees of hydrogenation. Different amorphous carbon grains are produced by condensation of carbon vapor obtained by laser pyrolysis of various hydrocarbons \citep{jager2006a} or laser ablation of carbon surfaces, or arc discharge between two carbon rods (\citet{mennella2003a} and references therein), or ion and photon irradiation of hydrocarbon-rich (e.g., pure $\mathrm{CH_4}$) ice layers \citep{dartois2005a, strazzulla1999a}. Carbon soots, produced by controlled combustion of acetylene or ethylene in a low-pressure flat flame burner, are of special interest because their precursors may be polycyclic aromatic hydrocarbons (PAHs) and fullerene-like molecules \citep{carpentier2012a}. These carbon dusts are mainly characterized by infrared \citep{gavilan2017a}, but also by Raman spectroscopy \citep{brunetto2009a}, which is sensitive to the carbon backbone structure and to its degree of order.

Laboratory experiments investigating how this ISM-inherited carbon dust could have been processed in the disk are of interest for cometary science. Once produced and characterized, these carbon dusts can be submitted to energetic processing encountered in space. Galactic cosmic ray and solar energetic particle irradiations are simulated by ion irradiation experiments, causing their progressive amorphisation \citep{brunetto2009a, mennella2003a, pino2019a}. The impact of these irradiations on PAHs or amorphous/hydrogenated carbon dusts embedded in ices is also of interest to understand if these carbon networks could break, incorporate heteroatoms (O, N, S), and link via aliphatic chains to form macromolecular network similar to IOM. Up to now, only few laboratory experiments have started to address this question, showing oxygenation/hydroxylation of PAHs molecules embedded in water ice after UV or proton irradiation \citep{bernstein2002b, yang2014a}.

\subsubsubsection{2.1.2.4. Gas phase organic synthesis}
Another way proposed to form pre-cometary organic dust is from gas phase or gas-solid phase chemistry in the hot ($\sim$1000 K) innermost region close to the protostar, followed by radial outward transport before incorporation into cometesimals. 
Gas-mineral Haber-Bosch (HB) catalytic reduction of $\mathrm{N_2}$ by hydrogen to make $\mathrm{NH_3}$, and Fischer-Tropsch type (FTT) catalytic reduction of CO by hydrogen in the presence of $\mathrm{NH_3}$, were early proposed as plausible sources of organic dust. Laboratory experiments have demonstrated their feasibility under nebular conditions, on various mineral grains (amorphous silicates, pure silica smokes etc.) and at low pressure (\citet{nuth2008a} and references therein; \citet{nuth2016}). These experiments take place in a glass apparatus where the mineral powder is subjected to a flow of gas at temperatures from 500 to 900 K (Figure \ref{fig:organicgas}). A carbonaceous coating is formed on the grains, and this coating is itself an efficient catalytic surface for the reaction to continue.

More recently, other experiments have been conducted to simulate radical-rich plasma environments of the protoplanetary disk (inner and irradiated upper layers of the disk, around 1 to 10 AU from the proto-star), by ionizing gas-phase mixtures \citep{bekaert2018a, biron2015a, kuga2015a}. These experimental setups, coined “\textit{Nebulotron}”, consist of a glass line in which a gas mixture ($\mathrm{CO}$, $\mathrm{N_2}$, $\mathrm{H_2}$) at about 1 mbar is flowed through a microwave generator. The microwaves trigger a plasma discharge in which gases are dissociated and ionized at 800-1000 K. Ions and/or radicals condense progressively on the glass surface, forming a refractory organic solid, which is analyzed \textit{ex situ}. This material reproduces some features of the chondritic IOM: hydrocarbon backbone structure \citep{biron2015a}, and elementary and isotopic signatures of noble gases \citep{kuga2015a}.

\subsubsubsection{2.1.2.5. Hydrothermal organic synthesis}
Refractory organics are also produced via condensation and polymerization reactions of small molecules in liquid water. Such reactions may occur in the interior of sufficiently large comets ($\geq$ 50 km, \citet{gounelle2008}), or asteroids. Moreover, because cometesimals have accreted some materials such as crystalline silicates that could have been formed in the innermost region of the protoplanetary disk, they might also have accreted some organic matter produced from hydrothermal reactions in parent bodies which were destroyed and whose materials were transported to cometesimals formation region \citep{cody2011a}.

Several experiments have shown that insoluble refractory organics, as well as soluble ones, can be synthesized in alkaline liquid water at $\sim$90-250°C from formaldehyde ($\mathrm{H_2CO}$) and ammonia ($\mathrm{NH_3}$) \citep{isono2019a, kebukawa2017a, kebukawa2013a}, but also from HMT \citep{vinogradoff2018a} (see section 2.1.2.2). The reactions leading to these compounds likely include formose and Maillard-type reactions (\citet{vinogradoff2018a} and references therein). Further works could address how various minerals can influence this hydrothermal chemistry \citep{vinogradoff2020a}, and provide estimates of how much of these organic solids could have been produced and incorporated in cometesimals. 

%--------------------------------- Figure Hadamcik2007 - Smoke
\begin{figure}[ht]
\includegraphics[width=\columnwidth]{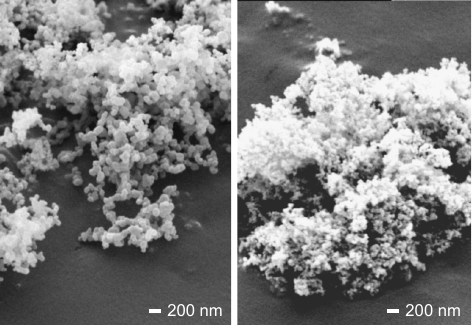}
\caption{SEM images for $\mathrm{MgSiO}$ (left) and $\mathrm{FeSiO}$ (right) smokes produced by gas-phase condensation \citep{hadamcik2007a}}
\label{fig:smoke}
\end{figure}
%---------------------------------

\subsubsubsection{2.1.2.6. Minerals}
Silicates observed in the diffuse interstellar medium are mostly amorphous \citep{do-duy2020a}, but comets contain a significant portion of crystalline silicates and refractory minerals (see the chapter by \textit{Engrand et al.} in this volume). These crystalline or refractory minerals have been formed at high temperatures, most probably in inner regions of the proto-planetary disk, before being transported and incorporated in cometesimals \citep{wooden2008a}. Open questions are how they formed, where they came from, when and how they were transported (see \textit{Engrand et al.} in this volume). 
Laboratory experiments have been conducted to study the formation of amorphous silicates, and their thermal processing to form crystalline silicates, as reviewed by \citet{colangeli2003a} and \citet{jaeger2009a}.

\textit{Production.} analogues of cosmic/cometary silicates are produced by gas-phase condensation, in a furnace at 500-1500 K containing atmospheres of $\mathrm{SiO}$-$\mathrm{H_2}$-$\mathrm{O_2}$ or $\mathrm{Mg/Fe}$-$\mathrm{SiO}$-$\mathrm{H_2}$-$\mathrm{O_2}$ \citep{nuth2000a, nuth2002a} at $\sim$100 or $\mathrm{10^-{}^7}$ mbar \citep{nagahara2009a}, or via laser vaporization/ablation of a silicate pellet, also called pulsed laser deposition \citep{brucato2002a, sabri2013a}. In these experiments, after vaporization, the cooling of the gas phase induces the nucleation of clusters growing into macroscopic solid grains, which are deposited on a substrate for \textit{in situ} (or collected, for \textit{ex situ}) analyses. Analogues of cometary silicates are also produced from sol-gel inorganic or organic chemical reactions, where a colloidal solution evolves towards a solid gel. After drying of the solution, amorphous silicates are obtained (\citet{thompson2019a} and references therein). Once produced, their composition and structure are characterized by a variety of usual techniques: Scanning or Transmission electron microscopy (SEM, TEM), Energy Dispersive X-ray (EDX), X–ray diffraction (XRD) etc., as detailed in section 7 of \citet{colangeli2003a}. The silicates produced are individual spherical grains 10 to 100 nm in diameter, arranged in fluffy chain-like agglomerates (Figure \ref{fig:smoke}), which are usually amorphous. Experiments producing silicates are constantly evolving with the aim to improve the production speed \citep{thompson2019a}, the reproducibility, the homogeneity of the dust, and to control experimental parameters such as oxygen fugacity (i.e., the amount of oxygen in the gas phase) representative of different pre- or protoplanetary disk environments \citep{wooden2017a}.

\textit{Post-formation processing.} After production, the amorphous silicates can be subjected to energetic processes simulating those encountered in the ISM and in protoplanetary disks, before their incorporation in cometesimals. Annealing is the process by which heating of amorphous silicates (for example after shocks in protoplanerary disks) results in the diffusion and rearrangement of structural units to form higher order structures, i.e. crystallization. Laboratory experiments are performed to determine the temperature at which crystallization can be triggered, and the kinetic of the annealing process (determination of the activation energy of crystallization) \citep{colangeli2003a, jaeger2009a}. In the ISM, the silicate dust is irradiated by cosmic ray ions, which lead to sputtering (erosion), amorphization and possible implantation of protons ($\mathrm{H^+}$). The annealing properties of such ion-irradiated silicate grains may be different from the original grains. Moreover, $\mathrm{H^+}$ implantation may form hydroxyl OH groups in the silicates before their incorporation in cometesimals, as shown by laboratory experiments \citep{djouadi2011a, jin2021a, mennella2020}. More laboratory experiments are needed to determine the ionization rate, implantation efficiency, and diffusion rate of hydrogen and deuterium in silicate minerals \citep{jin2021a}.

Future experiments should also address formation and evolution of other refractory minerals present in comets, such as $\mathrm{FeS}$, $\mathrm{Fe}$-$\mathrm{Ni}$ alloys, iron oxides and carbonates (see the chapter by \textit{Engrand et al.} in this volume). For example, experiments have shown that Mg-carbonates are produced with amorphous silicates during the non-equilibrium condensation of a silicate gas in a $\mathrm{H_2O}$-$\mathrm{CO_2}$-rich vapor \citep{toppani2005a}. Furthermore, there are tentative hints for the presence of small amounts of aqueously altered minerals in comets (\citet{rubin2020} and references therein; \textit{Engrand et al.} in this volume). Some of them might have been accreted from dust of aqueously altered bodies formed in the inner solar system and transported into cometesimals, or they have been produced in comets \citep{suttle2020a, gounelle2008}. Laboratory experiments on these aqueous alteration processes are described in section 2.3.4.2.

\subsubsection{Physical formation of a cometesimal, from grains to a cometary nucleus}

Laboratory experiments with the aim to study the growth of planets and other objects in the solar system have been extensively carried out in the last few decades. The experiments can be classified by the material used, or by the process to be studied. While experiments with non-volatile materials are easier to perform, experiments with water ice and other super- or hyper-volatiles are rare because of the complex handling of the icy materials. The growth of micrometer-sized particles to larger, mm-sized objects can be studied in so-called coagulation experiments. The further growth and evolution of the pebbles that were formed by the coagulation process can be studied in collision experiments to learn about the first growth phases in the young solar system.

\textit{Grains production and dispersion.} One major problem that occurs when performing laboratory experiments with micrometer-sized particles is that the sample material has to be dispersed prior to the execution of the experiments. This is often done by a particle dispenser. Depending on the material type, different methods for the dispersion process can be used. The best solution to create non-volatile micrometer-sized monomers is to fill the desired material (often silica, $\mathrm{SiO_2}$, is used) into a piston and the open end is pressed against a rotating cogwheel that disperses the material into monomers \citep{Blumetal2006}. Then, the produced particles are carried to the desired location by a gas stream. Often, a siphon is introduced into the pipe system to ensure that aggregates cannot follow the gas stream, and only single dust particles are collected on a filter at the output of the dispenser.

\begin{figure}[ht]
\includegraphics[width=\columnwidth]{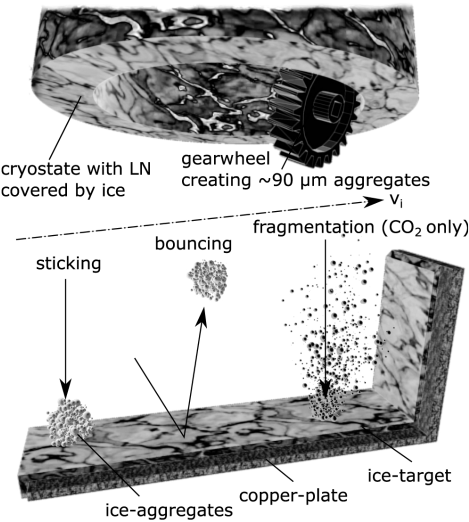}
\caption{Setup to study the collision of $\mathrm{H_2O}$/$\mathrm{CO_2}$ ice aggregates under 80 K and 0.1 to 1 mbar. Depending on the impact velocity ($\mathrm{v_i}$) of the aggregates colliding with the ice layer, sticking, bouncing or fragmentation is observed \citep{Musioliketal2016}.}
\label{fig:musiolik16}
\end{figure}

To create micrometer-sized water ice particles another method is required. Typically, a droplet dispenser is used to create a fog of micrometer-sized water droplets which are then forced into liquid nitrogen by a dry, cold gas stream \citep{Gundlachetal2011b, jost2013}. After evaporation of the liquid nitrogen, the sample material is accessible. For using water ice particles in dynamical collision experiments, the fog can be directly introduced into a cold pipe system. The droplets then freeze during flight through the pipes and are directed into the vacuum chamber \citep{GundlachBlum2015}. The creation of $\mathrm{CO_2}$-ice particles can be realised by adiabatic expansion of compressed $\mathrm{CO_2}$ gas. The expansion process cools the gas and $\mathrm{CO_2}$ particles are then created with typical grain size of $2 \, \mathrm{\mu m}$. Another technique to create larger $\mathrm{CO_2}$ particles/aggregates is to first create a solid layer of $\mathrm{CO_2}$ ice onto a cold surface which is then mechanically scraped by a rotating gearwheel (see Figure \ref{fig:musiolik16}) \citep{Musioliketal2016}. This method produces larger particles with a mean radius of about $90 \, \mathrm{\mu m}$, intra-mixtures can be realised by using several gas species for the same sample creation on the cold target, for example $\mathrm{CO_2}$ in combination with $\mathrm{H_2O}$ gas. 

Many experiments have been performed to study the first stage of planetesimal/cometesimal formation and the main findings are summarized in the following paragraphs.

\textit{Specific surface energy}. The specific surface energy is an important material parameter that describes the outcome of particle-particle collisions, as well as the strength of the cohesion of materials. It is therefore an important physical property that has to be known to simulate coagulation scenarios, collisional evolution of objects as well as cometary activity. Different types of experiments were performed to measure this physical property.

The first method was only used on $\mathrm{SiO_2}$ dust particles with radii ranging from $0.5 - 2.5 \, \mathrm{\mu m}$, but provides a precise measurement for the specific surface energy. Therefore, the adhesion force between micrometer-sized grains was determined by gluing single spherical $\mathrm{SiO_2}$ particles to the cantilever of an Atomic Force Microscope (AFM) and on the respective substrate \citep{Heimetal1999}. Then, the particles were brought into contact and were separated afterwards. To derive the specific surface energy from the force curves, the Johnson, Kendall and Roberts (JKR) theory derived by \citet{Johnsonetal1971} was used, which yielded a specific surface energy $\gamma = 0.019 \, \mathrm{J \, m^{-2}}$ for $\mathrm{SiO_2}$.

Another technique to measure the specific surface energy was introduced by \citet{BlumWurm2000}. The idea was to observe coagulating micrometer-sized particles when they formed a growing fractal agglomerate on a thin needle. At some point, parts of the grown agglomerate became too heavy and gravitational restructuring of these parts occurred. These restructuring events were observed with a camera to determine the mass of the restructuring part of the agglomerate. A determination of the specific surface energy was possible if the restructuring event could be described by rolling of the aggregate part around one particle. Under the assumption that the rolling friction force equals the gravitational force, these events were used to calculate specific surface energies for $\mathrm{SiO_2}$ \citep[$\gamma = 0.02 \, \mathrm{J \, m^{-2}}$;][]{BlumWurm2000} and for $\mathrm{H_2O}$ \citep[$\gamma = 0.19 \, \mathrm{J \, m^{-2}}$;][]{Gundlachetal2011b}.

The experiments discussed above indicate that the specific surface energy of water ice is an order of magnitude higher than the value for $\mathrm{SiO_2}$ and this result is not in agreement with the idea that $\mathrm{SiO_2}$ particles possess a surrounding water layer under terrestrial atmospheric conditions and also in vacuum experiments \citep{Kimuraetal2015}. Thus, it is expected that $\mathrm{SiO_2}$ should have the same specific surface energy as water ice. A possible explanation for the deviation seen in the experiments discussed above is that the water ice was not kept cold enough during these experiments. Recent studies have investigated the specific surface energy indirectly by measuring the tensile strength of cylindrical samples composed of either $\mathrm{SiO_2}$, or micrometer-sized water ice particles \citep{Gundlachetal2018}. The advantage of these experimental studies was that the environmental temperature was controlled, so that water ice temperatures below 140 K have been secured. These experiments have shown that water ice at low temperatures possesses the same specific surface energy compared to $\mathrm{SiO_2}$.

\textit{Sticking threshold velocity}. Direct collision experiments have been utilised to determine the sticking threshold velocity of the particles \citep{Poppeetal2000, GundlachBlum2015}. A particle jet was created and directed onto a quartz target, under vacuum conditions. The interaction of the particle jet with the target was studied with a long-distance microscope and a laser, which illuminated the particle jet in opposition. To avoid that direct laser light reached the long-distance microscope, a small absorption plate was mounted directly in front of the imaging system. With this setup, only the forward scattered light of the particles was detected by the microscope. Because the laser light was turned on and off with a very high frequency (stroboscopic laser light) the flight direction and the speed of the particles were measured. \citet{Poppeetal2000} used this technique to measure the critical velocity at which the particles stop to stick to the surface, the so-called sticking threshold velocity. At higher speeds, the particles are not able to stick to the surface and bounce off. \citet{GundlachBlum2015} refined this technique for use with water ice. The produced micrometer-sized ice particles were transferred into the vacuum chamber by a pipe system and the collision occurred with a cold ice target. Both works have shown that the sticking threshold velocity of water ice at low temperature (below $\sim$210 K) is ten times higher compared to the value for $\mathrm{SiO_2}$: $v_{St, H_2O} = 9.6 \, \mathrm{m \, s^{-1}}$ and $v_{St, SiO_2} = 1.0 \, \mathrm{m \, s^{-1}}$. At higher ice temperatures, the sticking threshold increases up to $\sim 50 \, \mathrm{m \, s^{-1}}$ which can be explained by the formation of a liquid-like layer on the surface of the grains \citep{Gaertneretal2017}, which could be the cause for the enhanced specific surface energy value measured by \citet{Gundlachetal2011b}.

\citet{Musioliketal2016} performed an additional set of experiments to study the collisional properties of $\mathrm{CO_2}$ ice as well as different ice mixtures. As shown in Figure \ref{fig:musiolik16}, the authors created different ice samples that were scratched off the surface by a rotating gearwheel. These particles were then attracted by Earth gravity and hit an ice target of the same composition underneath. The results of these experiments indicate lower values for the sticking threshold velocity, namely $0.04 \, \mathrm{m\,s^{-1}}$ for pure $\mathrm{CO_2}$ particles and $0.43 \, \mathrm{m\,s^{-1}}$ for a 1:1 mixture of $\mathrm{H_2O}$ and $\mathrm{CO_2}$.

%--------------------------------- Figure Musiolik16
%---------------------------------

\textit{Aggregation regimes.} \citet{BlumWurm2000} performed micro-gravity experiments to study the aggregation behavior of fractal agglomerates (fractal dimension of 1.9) under realistic conditions. For this purpose, a turbomolecular pump was modified to produce the fractal agglomerates during flight. The collisions of the formed agglomerates with a thin $\mathrm{Si_3N_4 }$ target occurred at velocities of $0.07 - 0.5 \, \mathrm{m\,s^{-1}}$ and were observed with a camera. The authors identified four aggregation regimes dependent on the impact velocity. Hit-and-stick collisions occur at the lowest velocities ($\sim0.2 \, \mathrm{m\,s^{-1}}$). Impact restructuring was observed for intermediate-low velocities ($\sim0.65 \, \mathrm{m\,s^{-1}}$), and at intermediate-high collision speeds ($\sim1.2 \, \mathrm{m\,s^{-1}}$) compact growth was observed. High-velocity ($\sim1.9 \, \mathrm{m\,s^{-1}}$) impacts led to fragmentation events. These experimental findings have confirmed the model developed by \citet{DominikThielens1997}.

\textit{The outcome of pebble collisions.} The collisional growth of small particles to larger objects (agglomerates and pebbles) was extensively studied in many experiments in the past years and their results were used as input for a very detailed collision model \citep[see Figure 1 in][and references therein]{Blum2018} that describes the different regimes for agglomerate-agglomerate collisions in the protoplanetary disk. In general two factors determine the collisional outcome: (1) the size ratio, and (2) the speed of the collision partners. Sticking collisions (please note that different aggregation regimes exist; see above) occur at low speeds and if the size ratio is close to unity \citep[see, e.g.,][]{WurmBlum1998, BlumWurm2000, Kotheetal2013}. Bouncing collisions take place when the energy dissipation during collision is insufficient to allow sticking \citep{Weidlingetal2012,Brissetetal2016,Brissetetal2017} and causes compaction of the pebbles \citep{Weidlingetal2009}. Fragmentation always occurs at larger speeds if the mass ratio is about unity \citep{Beitzetal2011,Schraepleretal2012,DeckersTeiser2013,Bukharietal2017}.

In case of different aggregate sizes, other collision results can occur. Mass transfer describes the process by which the smaller impacting aggregate loses mass by fragmentation during the collision \citep[see, e.g.,][]{Wurmetal2005b,TeiserWurm2009,Guettleretal2010,Beitzetal2011,DeckersTeiser2014,Bukharietal2017}. Typically 1 \% - 50 \% of the mass are transferred to the larger object. Cratering may occur if the projectile aggregate is larger than in the previous case. Then the target loses material, because more material is excavated than transferred \citep{Wurmetal2005a, Paraskovetal2007}. Cratering is a transition process to the fragmentation regime. Erosion is another process that can occur when very small ($< 0.1 \, \mathrm{mm}$) aggregates collide with a larger object at higher speeds ($>50 \, \mathrm{m\,s^{-1}}$) \citep{Bukharietal2017}.

All these experiments can be used to improve simulations that study the coagulation and collisional evolution in the protoplanetary disk in the early stages of planet formation \citep[see, e.g.,][]{Windmarketal2012a, Windmarketal2012b, Zsometal2010}. The collision behaviour of larger objects (from millimeter to meter) have been studied in laboratory experiments reviewed in \citet{guttler2013}. These experiments measured the coefficient of restitution (i.e. the ratio of the final to initial relative speed between two colliding objects) for spherical particles made of various materials (including water ice) up to meter size \citep{durda2011}.

Once formed, cometesimals and comets can experience collisions. Hydrodynamical models are used to predict the outcome of the collisions in terms of the general structure of the object and material properties such as porosity as well as their thermal history \citep{jutzi2020, schwartz2018}. These models use various experimental data as inputs, for instance the so-called “\textit{crush-curves}” (the relation between porosity and pressure) or the thermal conductivity and diffusivity of analogue material \citep{Guettleretal2009, Guettleretal2010}. While these properties are well known for the various refractory components of putative cometary dust, ices and complex organic molecules (pure or mixed) are more difficult to study. Systematic measurements with such materials would be very beneficial for future modeling studies. In the meantime, it is mandatory to derive high-quality scaling laws to apply the experimental findings on small icy and organic-rich particles to larger objects representative of the later stages of the formation process.

%---------------------------------
\subsection{Experiments on comet evolution during residence in the outer solar system}
\label{sec:subsection1}

\subsubsection{Brief summary of the processes}
After formation, the structure and composition of comets continue to evolve due to several processes, mostly irradiation by energetic particles and heating (see a review in \citet{weissman2020a}, and the chapter by \textit{Guilbert-Lepoutre et al.} in this volume).

Through the long residence time of comets in the Oort cloud or Kuiper belt/Scattered disk, galactic cosmic rays (GCRs) and solar energetic particles (SEPs) can deposit enough energy to transform the cometary materials chemical and/or physical properties down to tens of meters \citep{gronoff2020a, maggiolo2020a}. The surface and subsurface temperature of comets may also increase by 10 K or more due to supernovae (down to 2 m) or stars passing through the Oort cloud (down to 50 m) \citep{stern2003a}. In the case of a sufficiently large comet with a high dust-to-ice mass ratio, the decomposition of short lived radio nuclides ($\mathrm{^2{}^6Al}$) could provide internal heating \citep{prialnik2004a, Prialniketal2008}. Collisions with other bodies can also heat the surface and interior of comets \citep{schwartz2018a}, changing their composition and structure. Finally, other mechanical processes such as accretion or erosion caused by the impact of interstellar grains (reviewed in \citet{mumma1993a}), can also modify the surface or subsurface of comets.

Laboratory simulations of these processes could provide clues to distinguish the physical and chemical characteristics which are primordial, from the results of later processes or secondary evolution. This is particularly important to interpret correctly the observations of a comet approaching the Sun for the first time, which is the goal of the Comet Interceptor mission (see the chapter by \textit{Snodgrass et al.} in this volume).

\subsubsection{Irradiations by energetic particles}

Cosmic rays are particles (mostly protons, but also electrons and more massive ions) accelerated at energies up to several GeV \citep{bringa2003a}. They can penetrate the surface of cometary nuclei down to tens of meters, the more energetic the deeper. When they go through a material, such energetic ions can move atomic nuclei (elastic collisions) or produce nuclear reactions and/or ionizations and excitations (inelastic collisions), inducing re-arrangements of the material's structure and chemical reactions. Moreover, secondary particles produced from a single incident particle can also lead to a cascade of nuclear and ionizing reactions in the material. Molecules can dissociate and new ones can form, which are more or less volatile than the initial ones. These energetic projectiles can also cause sputtering, i.e. the expulsion of atoms or molecules from the surface, and changes of material structure (amorphisation, compaction). Finally, ions can be implanted down to a certain depth in the surface material and contribute to the formation of new molecules throughout that later.

\textit{Setups and products.} Laboratory experiments simulating these structural and chemical effects of GCRs usually use electrons, protons and heavier ions in the keV-MeV energy range, so with lower energy but higher particle fluxes than GCRs, to bombard a sample which is most often very thin (nanometers to micrometers thick) compared to the penetration depth of the radiation. The projectiles are generated by sources, such as electron guns (up to keV), ion accelerators including cyclotrons (up to MeV-GeV) or synchrotrons (up to GeV). Varying the mass and the energy of the ions allows for the investigation of either elastic or inelastic interactions and for varying the stopping power (i.e., the energy loss during collision) of the projectile in the material. From these experiments with multiple projectiles, the relation between the sputtering or radiolysis cross sections and the stopping power can be calculated. Knowing the flux of GCRs as a function of their energy in the Oort cloud or Kuiper belt, the temporal evolution of cometary materials (via sputtering, radiolysis, amorphization etc.) can be estimated (see \citet{rothard2017a} for details). Sophisticated numerical models taking into account many parameters (generation of secondary species by projectiles, diffusion of products etc.) can estimate how deep the nucleus material is affected by these processes \citep{gronoff2020a, maggiolo2020a}.

\textit{Ion irradiation on ices.} In section 2.1.2, we described the laboratory experiments studying the chemical effects of photolysis and radiolysis in pre-cometary ices. The same chemistry continues to occur in cometary ices after comet formation. \citet{rothard2017a} and \citet{allodi2013a} have provided detailed reviews of the laboratory experiments simulating the modifications of ices caused by cosmic rays and solar wind. A large number of ion or electron irradiation experiments have been performed on single-component ices ($\mathrm{H_2O}$, $\mathrm{CH_4}$, $\mathrm{N_2}$ etc.), $\mathrm{H_2O}$-rich ices ($\mathrm{H_2O}$ being the dominant ice in comets), or $\mathrm{N_2}$-rich ices (because $\mathrm{N_2}$ is largely present on trans-neptunian objects TNOs, and so possibly on distant comets), as reviewed in \citet{hudson2008a}. More recently, experiments on methanol ($\mathrm{CH_3OH}$)-rich ices have also been performed \citep{urso2020a}. Experiments performed with doses in the range of about 1–20 eV/16amu allow the monitoring of the progressive formation of new molecules, analyzed mostly via infrared spectroscopy and mass spectrometry as described previously in section 2.1.2.1 (Figure \ref{fig:icechem}). But at higher doses of the order of 100 eV/16amu, carbon-containing ices (such as $\mathrm{CH_4}$, $\mathrm{CH_3OH}$, or $\mathrm{C_6H_6}$) progressively lose $\mathrm{H_2}$ and increase their carbon-to-hydrogen ratio: the produced molecules bind together, forming colorized refractory organics and finally an organic crust, masking the ice below it \citep{brunetto2006a, strazzulla2003a, strazzulla1991a}. The colorization of ices by ion irradiation is especially studied in the case of TNOs \citep{dalle2011a}. Experiments have shown that the total deposited energy (elastic plus inelastic collisions) is the most important parameter in the reddening process \citep{brunetto2006a}. Such laboratory experiments also tested the hypothesis of the formation of the N-rich organic matter observed in Ultracarbonaceous Antarctic Micrometeorites (UCAMMs) by GCR irradiation of $\mathrm{N_2}$-$\mathrm{CH_4}$-rich ices at the surface of a cometary body in the Oort cloud \citep{aug2016a}. At high irradiation dose, these organic residues evolve toward a hydrogenated amorphous carbon \citep{baratta2008a}. The carbonization process is efficient only for bodies accumulating a sufficient dose of GCRs such as comets in the Scattered disk or in the Oort cloud, where this crust of highly processed organic material may extend down to several meters \citep{strazzulla2003a, strazzulla1991a}. GCRs may change not only the chemistry, but also the structure of the ices in comets inducing amorphization of crystalline ices, and compaction of amorphous ice with time. These phenomena have been described by experiments performed in the context of interstellar water ice mantles \citep{dartois2015a}, or icy satellites surfaces \citep{strazzulla2013a}, but they may apply in the context of comets. In comets, the amorphization induced by GCRs may compete with the crystallization induced by thermal waves (see sections 2.2.3 and 2.3.4.1). 

\textit{Ion irradiation on refractories.} Studies were also dedicated to the effects induced by ion irradiation on refractory materials mainly representative of the Moon and of asteroids, but some of these results are relevant to comets as well. Ion irradiation of minerals causes their amorphization, and the formation of nano-phase reduced iron (np-$\mathrm{Fe^0}$) if they are Fe-bearing minerals \citep[for reviews, see ][]{bennett2013a, brunetto2015a}. These nanoparticles cause the darkening and spectral reddening of irradiated silicates \citep{hapke2001a}. However, complex mineral mixtures, such as carbonaceous chondrites, are bluing or reddening upon ion irradiation \citep{lantz2017a}. Ion irradiation was also performed on materials whose chemical structures can be considered analogous to some cometary refractory carbonaceous compounds, such as natural hydrocarbons (asphaltite, kerite) \citep{baratta2008a, moroz2004a} and soots produced in a low pressure flame \citep{brunetto2009a}. These experiments have shown the progressive carbonization of the samples, evolving toward amorphous phases. Irradiated soots exhibit Raman spectra similar to those of some meteorites, IDPs, and Comet Wild 2 grains collected by the Stardust mission \citep{brunetto2009a}. Visible and near-infrared reflectance spectra of the natural hydrocarbons evolve from red to neutral after ion irradiation \citep{moroz2004a}. These changes were mainly caused by elastic collisions, in contrast with ices whose evolution is controlled by the total of elastic and inelastic collisions.

\textit{Toward experiments with thicker and mixed samples.} Most of these experiments have been done on nanometer to micrometer thick samples of ice or dust. However, highly penetrating energetic particles interact with a much thicker layer of cometary materials (up to several meters) having some porosity and being a mixture of ice and dust components. These large-scale structural and compositional properties certainly induce effects that are not apparent in small-scale irradiation experiments performed up to now. These effects can be heating, charging, and diffusion of radiolysis products on the surface or in the volume of the ice, inducing new/enhanced chemical reactions and structural changes. For example, experiments performed on thick and porous water ice samples have revealed the charging and sublimation of the ice upon irradiation by ions \citep{galli2018a}.

\subsubsection{Thermal evolution}

The heat experienced by cometary nuclei during their residence in the outer solar system, due to the decay of short lived radio nuclides ($\mathrm{{}^2{}^6Al}$) or due to the influence of nearby stars or supernovae or to impacts, can change the composition and structure of a significant fraction of their surface and interior \citep{stern2003a, prialnik2004a, Prialniketal2008, schwartz2018a, mumma1993a}. Moreover, when comets approach the Sun for the first time, their surface is heated by the solar radiation. Due to the very low thermal conductivity of granular materials at their surface, the heat cannot efficiently penetrate into the interior of the nucleus. This also implies that cometary nuclei should experience alteration of their orbital elements due to the Yarkowski effect and rotational spin up by the Yarkovsky–O'Keefe–Radzievskii–Paddack (YORP) effect \citep{Bottkeetal2006}. Both effects strongly depend on the thermal conductivity of the cometary materials.

Laboratory experiments studying these thermal properties and processes are described in the following section 2.3, for comets penetrating the inner solar system. Thermal processes are exacerbated during the approach to the Sun but some of them may also occur to some extent during/after the formation of comets and during their residence in the Scattered disk or Oort cloud. Experimentalists should keep in mind that these thermal properties and processes are strongly influenced by the differences of composition and structure of the body considered, inherited from its history, and whether this body is a cometesimal, a pristine or slighly heated comet (depending on its mass and formation history), or a comet significantly more altered after multiple passages close to the Sun. Future experiments should consider each of these cases, and clearly differentiate between them, to improve our understanding of the thermal evolution of these bodies.

%---------------------------------
\subsection{Experiments on comet evolution during approach to the Sun}
\label{sec:subsection2}

\subsubsection{Brief summary of cometary activity processes}
When a comet penetrates the inner solar system, the absorption of the incident solar light by the surface dust and/or ice components causes the cometary surface to heat. The heat is then transported to deeper layers of the nucleus (see the chapter by \textit{Guilbert-Lepoutre et al.} in this volume). Cometary nuclei being extremely porous, we can distinguish three mechanisms of heat transport: (1) the conduction through the solid network of particles, (2) the radiative heat transport originating from thermal radiation through void spaces between particles, and (3) the gaseous heat transport due to the gas present in the void spaces. The effective thermal conductivity of porous materials is then given by the combination of these three effects. The heat wave can cause various physical and chemical transformations: phase-changes and segregation of the solid ices or salts, sublimation, release of trapped gases, decomposition of clathrates, and activation of chemical reactions. In return, these transformations can release or absorb energy inside the nucleus. For instance, the phase transition of pure water ice from amorphous to crystalline releases a significant amount of energy but the same process can be endothermic if the ice contains impurities \citep{kouchi2001a}. Moreover, the inward flowing gas from the sublimation front to the nucleus interior leads to the recondensation of ices in deeper and cooler layers, acting as a transport of latent heat of condensation toward the interior (see Figure \ref{fig:rickman}b). Layers made of components having similar volatility can thus form in the nucleus, leading to what is called nucleus chemical differentiation. These layers could get reduced porosity. Chemical transformations (isotopic fractionation, gas-dust reactions etc.) could also occur during diffusion. At the surface, the sublimation of the ices leads to the formation of a dessicated surface layer made of the most refractory components. Gases flowing from the sublimation fronts at depth to the surface can break fragments of this layer, or of sub-surface layers, and eject dust/ice particles in the coma.

All these processes are responsible for not only shaping the nucleus interior and surface, but also the coma. Indeed, a good knowledge of the mechanisms producing the gases and ejecting the ice/dust particles is essential for a correct interpretation of the observations of outgassing comets. Furthermore, the planning of any future cometary nucleus material sampling for \textit{in situ} analysis or sample-return will require a good knowledge of the (sub)-surface nucleus properties. Numerical models taking into account most of these multiple processes have been developed to simulate the nucleus and coma evolution of periodic comets \citep{marboeuf2012, Prialniketal2008}. Experiments carried out to understand the different steps of this energy flow, and the numerous parameters and processes controlling these physical and chemical transformations of comets, are described in the paragraphs below.

%--------------------------------- Figure Haack2021 - Sublimation
\begin{figure*}
\begin{center}
\includegraphics[width=17.1cm]{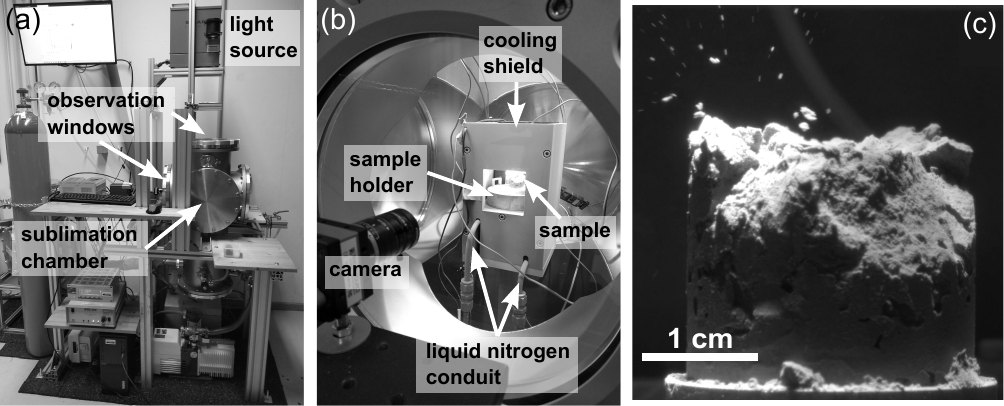}
\caption{Setup for sublimation experiments at the Technical University of Braunschweig. (a-b) External and internal views of the vacuum chamber equipped to monitor the sublimation of several-cm-large cometary analogue samples. (c) Ejections of icy-dust fragments during the sublimation of a sample made of water ice (20 vol\%) mixed with fly ash (mostly $\mathrm{SiO_2}$ and $\mathrm{Al_2O_3}$) particles \citep{haack_sublimation_2021}.}
\label{fig:haack21subli}
\end{center}
\end{figure*}
%---------------------------------

\subsubsection{Large-scale insolation experiments}
With the aim to simulate experimentally the evolution of a cometary nucleus during its approach to the Sun, several experiments have studied the consequences of incident solar-like (or infrared) light illuminating the surface of complex cometary analogue samples made of porous granular mixtures of ices and non-volatile constituents. These experiments are performed in simulation chambers allowing to maintain such samples, generally several centimeters large and thick, under high vacuum (10$^-{}^5$-10$^-{}^6$ mbar) and cometary-like temperatures (100-250 K). Typically, a solar simulator irradiates the sample surface, and the transformations of the sample surface and interior are monitored using various instruments (temperature sensors, cameras, hyperspectral imaging system, penetrometer, scale etc.) (see Figure \ref{fig:haack21subli}a,b) . A mass spectrometer can also be used to monitor the released gases. In the 70's, at the Ioffe Institute in Leningrad and at the Institute of Astrophysics in Dushanbe, E. Kajmakov and co-workers prepared mixtures of ice and minerals and/or organics and observed the formation ‒and outbursts‒ of a porous ice-free matrix after sublimation of ices at 10$^-{}^5$-10$^-{}^6$ mbar and 180-240 K \citep{dobrovolsky1977}. In the 80's, at the Jet Propulsion Laboratory in Pasadena, \citet{saunders1986} studied the sublimation of water ice samples prepared from a suspension of various silicates and organic matter \citep{storrs1988}. They noticed that some classes of phyllosilicates and organics were able to form fluffy filamentary sublimate residues, whereas most classes of silicates were not. Later, in the 80's and 90's, the Kometen Simulation (KOSI) project at the Deutsches Zentrum für Luft- und Raumfahrt (DLR) conducted a total of 11 experimental simulations over 7 years, considerably enhancing our understanding of cometary activity \citep{sears1999a, grun1991, kochan1999}. Many different properties of the samples (made of $\mathrm{H_2O}$, $\mathrm{CO_2}$ ices and various silicates and organics) as well as the evolved gases were studied in detail. \citet{bar-nun2003a} constructed a simulation chamber in which large (200 cm$^2$ $\times$ 10 cm) samples made of fluffy agglomerates of 200-µm gas-laden amorphous ice grains were prepared at 80 K and 10$^-{}^5$ mbar. These samples were irradiated from above using infrared radiation in order to study the evolution of gases released, and of thermal and mechanical properties. More recently, the Comets Physics Laboratory (CoPhyLab) team constructed at the Technical University of Braunschweig a new laboratory hosting several small-scale experiments as well as a large-scale comet-simulation chamber (L-Chamber). The latter one, equipped with 14 different scientific instruments, allows uninterrupted experiments such as solar illumination cycles on 10-30 cm thick and 25 cm diameter dust-ice samples, at low temperatures and pressures for up to several weeks \citep{kreuzig2021}.
The main results of these large-scale insolation experiments are described in the paragraphs below. They allowed the study of the numerous structural and compositional transformations occuring after insolation (for examples, see Figures \ref{fig:haack21subli} and \ref{fig:rickman}a). However, to understand how each experimental parameter (initial sample properties, temperature etc.) influences these transformations, it is necessary to perform a series of experiments in which a single parameter is changed at a time. This was not always the case in past experiments, such as KOSI. Moreover, most of these experiments also used initial ice-dust samples having relatively low refractory-to-ice mass ratio compared to the range of values (from 0.3 to 7) estimated from the Rosetta mission instruments \citep{choukroun2020}. Future experiments of this kind should thus be more systematic and investigate samples having larger refractory-to-ice mass ratio. Other smaller-scale experiments, focused on peculiar processes or properties using simpler samples, are complementary to the large-scale ones, and equally essential to provide physical constants as input to numerical models.

\subsubsection{Experiments on heat transport}

In cometary nuclei, heat can be transported due to the conduction through the solid network, the radiation inside the pores, and the gas diffusion \citep{gundlach2012a}. Numerical models of comet nuclei thermal evolution need the input of numerous physical parameters including the thermal conductivity (in W.K$^-{}^1$.m$^-{}^1$), which determines the amount of heat transported from the insolated surface into the interior of cometary nuclei and thus the temperature gradient. In the laboratory, samples deposited on a cold plate can be irradiated by a laser or a light source. The temperature evolution of the illuminated surface and bottom surface of the sample can be measured using an infrared camera and temperature sensors, respectively. The thermal conductivity can then be derived from this temperature evolution \citep{krause2011a, gundlach2012a}. Such experiments can be performed under vacuum or not and the sample size can be of several centimeters or larger. Numerous experimental measurements of thermal conductivity of porous media have been performed to test models that allow the calculation of the thermal conductivity from a set of parameters such as the particle size and composition, porosity, and temperature \citep{gundlach2012a, sakatani2017a,wood2020a}. In a porous medium, the thermal conductivity can be reduced by several orders of magnitude compared to the conductivity of the bulk material. Indeed, due to the small contact areas between the particles, which determine the efficiency of the heat exchange between the particles, the thermal conductivity decreases with increasing porosity or with decreasing filling factor \citep{arakawa2017a, krause2011a, sakatani2017a}. The radiative thermal conductivity depends on the particle size and on the mean free path of the photons inside the pores \citep{gundlach2012a, sakatani2017a}. If the porous medium is not under vacuum, the thermal conductivity depends also of the static gas pressure inside the pores. \citet{seiferlin1996a} measured a matrix conductivity close to 0.02 W.m$^-{}^1$.K$^-{}^1$, but maximum values for the effective (= matrix + vapor) thermal conductivity exceeding 0.25 W.m$^-{}^1$.K$^-{}^1$ at large pressures close to 1 bar. We can also note that most of the experiments designed to measure the thermal conductivity of cometary analogue samples have used pure water ice, whereas the gas trapped in amorphous water ice can have an impact on the thermal conductivity of the water ice. Indeed the thermal conductivity of gas-laden ice is much higher than for pure ice, and increases with the concentration of trapped species \citep{bar-nun2003a}. This effect of trapped species in water ice on its thermal conductivity is poorly known and new experiments, with relevant concentration of trapped species, are desirable.

Moreover, to model the temperature gradient in cometary nuclei, one should also consider the inward flowing of gas from the sublimation front to the nucleus interior, leading to the recondensation of ices in the deeper and cooler interior (Figure \ref{fig:rickman}b). This inward flow of gas and the ice recondensation can be seen as a transport of latent heat towards the interior \citep{benkhoff1995a}. This latter mechanism can be significant. Indeed, an experiment performed on a pure, porous water ice sample allows to estimate that the energy transport by the inward flowing of gaseous water represents 40 \% of the total heat flux available for heating \citep{benkhoff1995b}.

The thermal conductivity of cometary nuclei is generally assumed to be homogeneous and constant in time. Nevertheless, some KOSI experiments have provided experimental evidence of evolution of the studied sample, such as the recondensation of ices and the formation of a dust mantle (Figure \ref{fig:rickman}a), which could lead to a variation of the thermal conductivity as a function of time and nucleus depth \citep{huebner2006a}. For example, the sintering of ice grains leads to a growth of the grain to grain contact area enhancing the thermal conductivity \citep{Kossacki1994a}. The thermal conductivity of the dust mantle seems to depend on the presence of organics. \citet{koemle1996a} studied the evolution of an originally homogeneous multi-component sample (containing water ice, organics, and minerals) to a residuum containing (finally) only minerals and organics. During this evolution the thermal properties changed dramatically. The heat conductivity of the cohesive residuum was found to be at least an order of magnitude larger than the typical value for a loose dust mantle containing no organic material.

While the effect of numerous parameters, such as the grain size and composition, porosity and temperature, on the thermal conductivity of porous samples are well known \citep{gundlach2012a, sakatani2017a, wood2020a}, most of the experiments aimed to measure the thermal conductivity of cometary analogues consider mixtures made of only minerals and water ice. The impact of refractory organics and ammonium salts, which could be important components of the cometary nuclei, on the thermal conductivity are almost unknown. In a similar way, if the thermal conductivity of pure water ice is well known, that of gas-laden amorphous water ice should be studied further.

\subsubsection{Experiments on ice thermal transformations}

\subsubsubsection{2.3.4.1. Thermally induced physical transformations}
\textit{Different forms of ices.} The ice found in comets is mainly water (70-90 \%), along with other minor species possibly mixed with water. Depending on its place of formation and evolution before it approaches the Sun, different types of water-rich ice mixtures may be found in a comet, apart from pure crystalline or amorphous water ice: mixtures of water ice with other compounds, or molecules forming strong hydrate complexes, clathrate hydrates, or a mixture of these different structures. Clathrate hydrates are crystalline solids where water molecules are organized in the form of individual cages, stabilized by hydrogen bonds, trapping a single molecule other than water. The thermodynamic properties of each of these structures are different, and need to be measured via experiments.

\textit{Crystallization.} The crystallization of amorphous water ice can occur during the heating of cometary nuclei. The rate of this phase transition has been experimentally measured by \citet{schmitt1989a}. If the amorphous water ice does not contain impurities, this process is exothermic and the heat release is about 9 $\times$ 10$^4$ J.kg$^-{}^1$ \citep{ghormley1968a}. Nevertheless, for high enough impurity content, the process is endothermic \citep{kouchi2001a}. The negative energy balance during crystallization may be caused by the energy lost in expelling the guest molecules from the water lattice. 

\textit{Sublimation.} The heat experienced by the ices contained in the cometary nuclei leads to the production of gases and cometary activity. The first step to model cometary activity is to consider that cometary nuclei contain only pure ices. The temperature dependence of the sublimation vapor pressure for pure ices has been reviewed by \citet{fray2009a} who have compiled more than 1800 experimental measurements of vapor pressure for 53 molecules and propose some vapor pressure relations which can be included in models. Nevertheless, the reality is much more complex and most of the cometary molecules could be trapped in water ice. Some volatile species, such as CO, N$_2$ or Ar, can be partially trapped during the condensation of amorphous water ice even if the condensation temperature is too high to allow the direct condensation of these volatile species. This process has been intensively studied by Bar-Nun and coworkers \citep{bar-nun1988a, bar-nun1998a, bar-nun2007a, greenberg2017a}.

\begin{figure}[ht]
\includegraphics[width=\columnwidth]{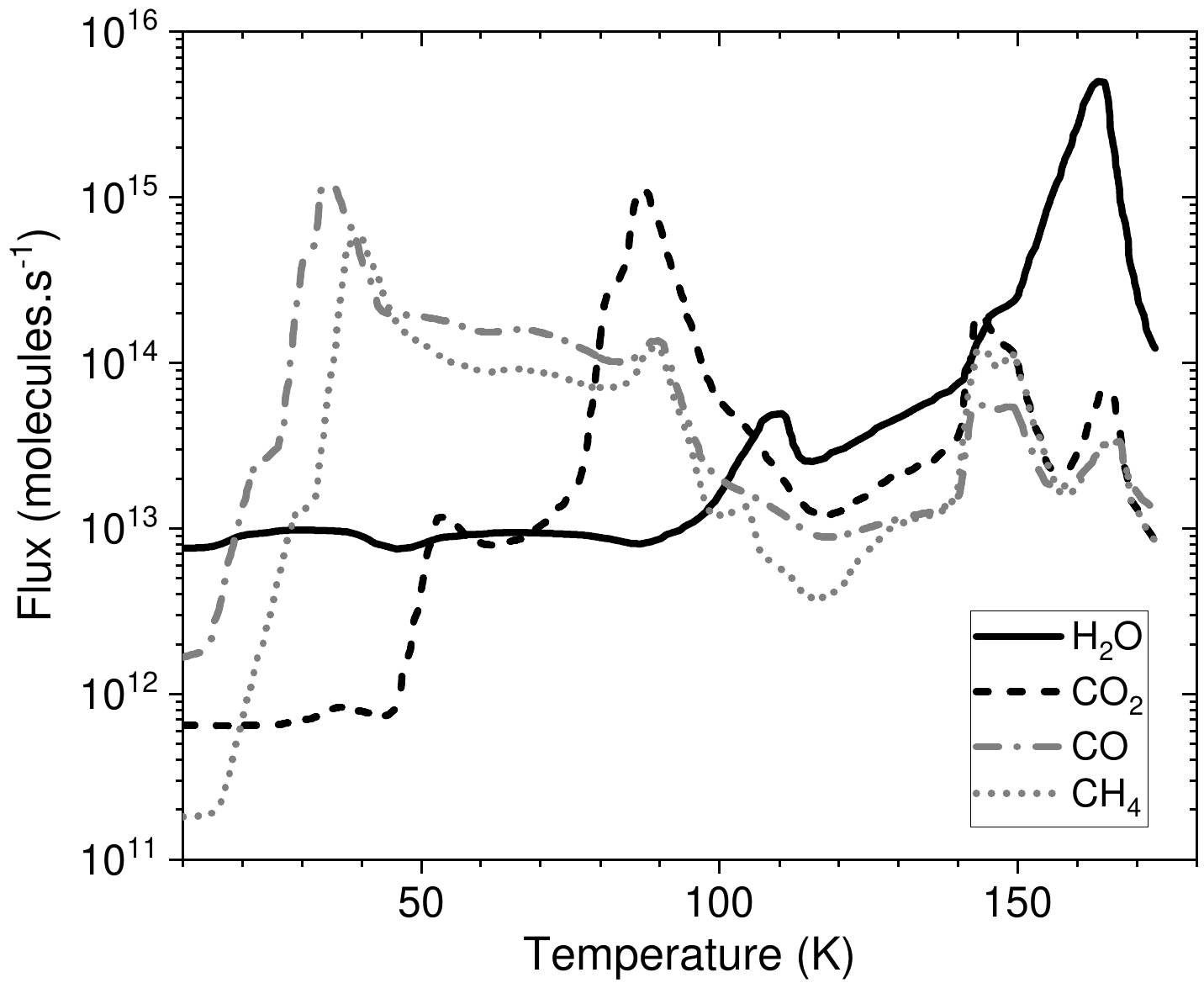}
\caption{Molecular fluxes of $\mathrm{CO}$, $\mathrm{CH_4}$, $\mathrm{CO_2}$ and $\mathrm{H_2O}$ as a function of the temperature from an ice mixture condensed at 10 K \citep{kouchi1995a}. The initial ratio at 10 K is $\mathrm{H_2O}$:$\mathrm{CO_2}$:$\mathrm{CH_4}$:$\mathrm{CO}$ = 65:10:10:15. The sublimation behavior is similar to that of a $\mathrm{H_2O}$-$\mathrm{X}$ binary mixture, but the very broad peak ranging from 20 K to 90 K is due to the release of $\mathrm{CO}$ and $\mathrm{CH_4}$ trapped in an amorphous $\mathrm{CO_2}$ ice. Such experiments show the necessity to study the sublimation from complex ice mixtures, and not only from $\mathrm{H_2O}$-$\mathrm{X}$ binary mixtures, to understand the processes taking place in cometary nuclei.}
\label{fig:kouchi95}
\end{figure}

\begin{figure*}
\begin{center}
\includegraphics[width=17.1cm]{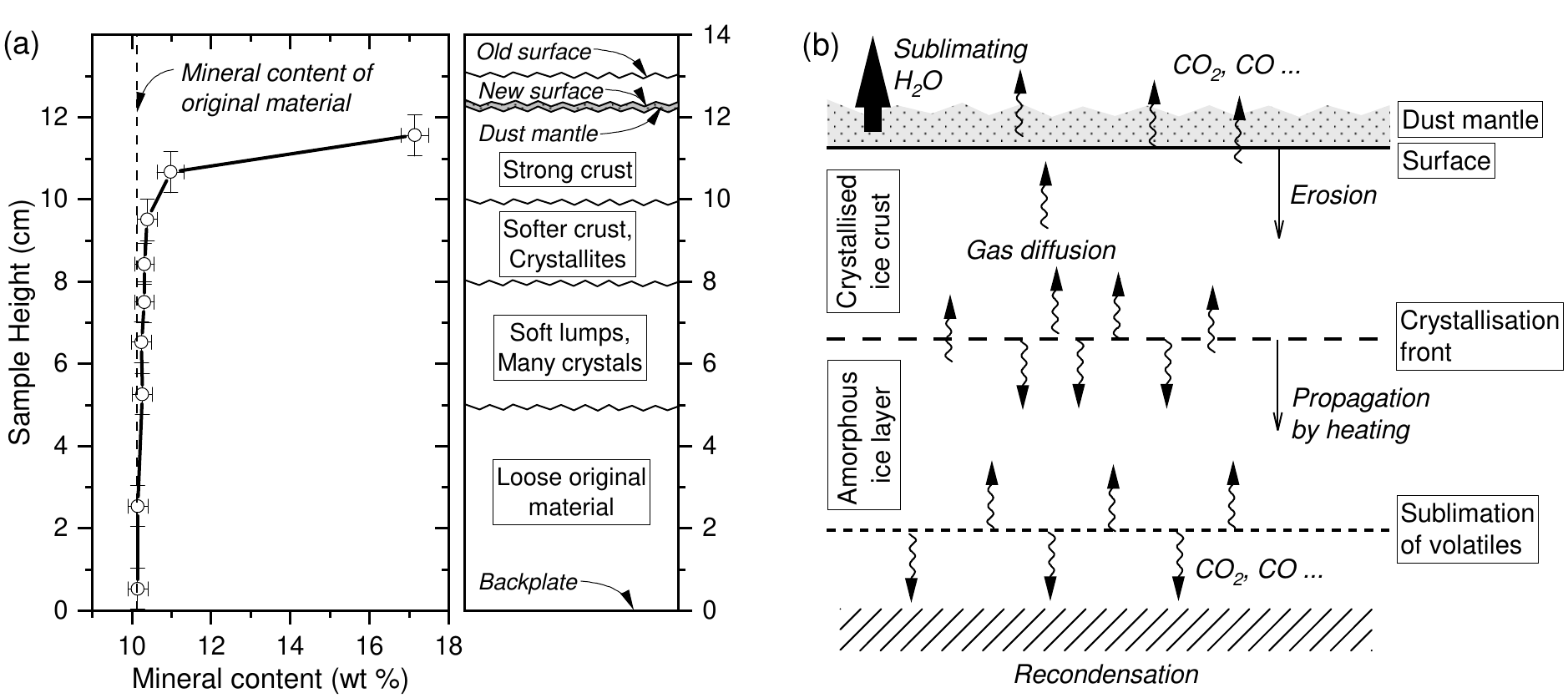}
\caption{Possible stratification in a comet nucleus approaching the Sun. (a) Cometary analogue sample mineral content and macroscopic stratigraphy after a large-scale insolation experiment (KOSI 9, \citet{gruen1993a}) with a sample made of 10\% dust and 90\% crystalline water ice (see section 2.3.2). (b) Possible processes in a short-period comet nucleus \citep{rickman1991, sears1999a}. The insolation of the surface is followed by heat propagation processes inside the nucleus (see section 2.3.1). At the surface, the sublimation of the ice causes erosion and the formation of a dust mantle. Deeper, the water ice initially amorphous (contrary to the experiment on the left) becomes crystalline. During crystallisation, volatile molecules initially trapped in water are released. Some gas molecules diffuse upward through warmer material, other diffuse inward and recondense into volatile-enriched layers.}
\label{fig:rickman}
\end{center}
\end{figure*}

The sublimation of binary mixtures has also been extensively studied. When water ice trapping a hyper-volatile species, such as CO, N$_2$ or Ar is slowly heated, the release of the more volatile components occurred in several discrete temperature regions. Thus, a fraction of the hyper-volatile species can remain trapped in water ice until its complete sublimation (see Figure \ref{fig:kouchi95}) \citep{bar-nun1985a, bar-nun1988a, bar-nun1998a, hudson1991a, kouchi1990a, kouchi1995a}. An extensive experimental study of the vaporization of binary mixtures has been performed by \citet{collings2004a}, allowing to separate the different species into three categories based on their desorption behavior. The sublimation of more complex ice mixtures, containing numerous species, has been studied by \citet{kouchi1995a, mart2014a, potapov2019b}. In that case, the desorption of the different species as a function of the temperature is rather complex and some effects caused by the presence of $\mathrm{CH_3OH}$ in the mixture can not be reproduced in experiments with binary mixtures. Such experiments show the complexity of the sublimation processes that take place in cometary nuclei.

When heated, the clathrate hydrates release the molecules that were trapped. The experimental data available on the equilibrium between clathrate hydrates, water ice and gas have been reviewed by \citet{Lunine1985a, Choukroun2013a}. For some species, like $\mathrm{O_2}$ or $\mathrm{N_2}$, experimental data are available only at pressure larger than 10 bar. To be applied to the cometary nuclei, these data must be extrapolated at lower pressures. As the empirical laws, such as $\mathrm{ln P = A + \frac{B}{T}}$ are generally applicable only on moderate pressures ranges, the validity of such extrapolation is questionable. For some other species, like $\mathrm{CO}$, there are no experimental data for the equilibrium at low pressure and temperature. Moreover, the available experimental data are relevant only for pure clathrate hydrates; the stability of mixed clathrate hydrates at low pressure and temperature is very poorly known. For all these reasons, new experimental measurements on the stability of clathrate hydrates would be welcome.

%--------------------------------- Figure Kouchi95
%---------------------------------

\subsubsubsection{2.3.4.2. Thermally induced chemical transformations} 

\textit{Thermally induced ice chemistry.} The volatile molecules constituting ices are primarily affected by heating processes, as explained in the previous section 2.3.4.1. The increase of ice temperature leads to the rearrangement of the molecules, from amorphous to crystalline phase, and to the separation of mixed molecules, i.e. segregation. Heating can also lead to the sublimation of molecules, passing from the solid to the gas phase. Simultaneously to these structural changes, volatile molecules initially trapped in the amorphous water ice can be released. This enhanced molecular mobility favors chemical reactions in the solid phase, and possibly the gas phase, inside cometary nuclei.

The dynamics of solid state chemical reactions at the surface and inside the ice grains is controlled by both the diffusion coefficients of the reactants and the reaction rate constants. Laboratory experiments are performed to measure these parameters serving as inputs to numerical models of chemical reactions networks (see review in \citet{theul2019a}). Although these experiments are primarily done in the context of pre-cometary interstellar grains, they are also relevant to the chemistry happening in cometary nuclei ice mixtures. Using setups similar to those described in section 2.1.2.1 and Figure \ref{fig:icechem}, these experiments are studying reactions between neutral molecules (neutral-neutral, such as condensation, acid-base, and nucleophilic addition, \citet{theul2013a}), or reactions with radicals formed from interactions with GCR/SEP (neutral-radical and radical-radical reactions, \citet{linnartz2015a}). Recent experiments have shown that diffusion of molecules on the surfaces of ice (external or internal surfaces such as cracks or pores) is much faster than inside the bulk ice \citep{minissale2019a, mispelaer2013a}. Structural changes induced by heating result in the formation of many cracks in the ice \citep{may2012a}, allowing the diffusion of reactants along them \citep{ghesqui2018a}. Barrier-less or low activation energy reactions then form many new molecules, resulting in the so-called “\textit{organic refractory dust}” (Figure \ref{fig:danger13}). Laboratory investigations of this organic dust were detailed in section 2.1.2.2. An open question is how much of the salts, polymers or macro-molecules constituting this organic dust are formed during the residence in cometary reservoirs and during the active phase of comets, compared to those which were incorporated in cometesimals. Another question to investigate is how these reactions, triggered by heating, modify the isotopic composition of cometary ices via exchanges of $\mathrm{D/H}$ atoms \citep{faure2015a, lamberts2015a}.

\textit{Liquid water-induced chemistry?} If temperature and pressure conditions allow the melting of the water ice in the subsurface of cometary nuclei, even transiently, this liquid water may induce additional chemistry, affecting mineral and organic phases. Analogues of pre-cometary amorphous silicates produced by gas-phase condensation experiments (see section 2.1.2.6) have been reacted with liquid water at different temperatures (331-378 K) and reaction times (1-142 h) \citep{nelson1987a, rietmeijer2004a}. The porous amorphous silicate particles were transformed into phyllosilicates, and it was observed that the swelling of phyllosilicates decreased the porosity of the dust, limiting further propagation of the liquid water, and leading to localized alteration \citep{rietmeijer2004a}. Other experiments conducted on CP-IDPs suspended in liquid water showed that their amorphous silicates are altered into phyllosilicates two orders of magnitude faster than laboratory analogues \citep{nakamura-messenger2011a}. Based on these studies, in addition to numerical modelling and observations, \citet{suttle2020a} proposed that for some comets passing within $<$ 1.5 AU of the Sun, the solar radiation would generate temperatures high enough in their immediate subsurface ($<$ 40 cm) for the water ice to melt, potentially forming the minor secondary minerals observed in some IDPs and comets. In addition to phyllosilicates, future laboratory experiments could also investigate the formation of other minerals, as well as the organic chemistry potentially produced by transient liquid water on comets.

%--------------------------------- Figure KOSI Rickman
%---------------------------------

\subsubsection{Experiments on gas diffusion from the nucleus interior to the surface and coma}

\subsubsubsection{2.3.5.1. Experiments on gas diffusion through the dust}

\textit{Stratification.} Large-scale cometary insolation experiments, especially KOSI, have shown that the propagation of the heat wave inside an initially homogeneous mixture of ice and dust particles results in the vertical stratification of the sample (see Figure \ref{fig:rickman}a). The loss of water and other volatiles from the surface leads to its erosion and to the formation of a refractory dust mantle on top of the sample. This mantle, also called sublimation residue, is often quite porous and loosely consolidated with a micro-structure of filaments made of aggregated dust grains \citep{saunders1986, poch2016b, sears1999a}. Below this dust mantle, a system of ice layers formed by chemical differentiation have been identified. The sublimation of ices generates gases that diffuse inward and outward of the sublimation front. These gases can then recondense on cold-enough regions (recondensation fronts), especially at greater depth. After most of the KOSI experiments, a consolidated layer formed by recrystallization of $\mathrm{H_2O}$ and $\mathrm{CO_2}$ was found just below the dust mantle, and 200-µm pure ice crystals were also observed \citep{hsiung1989, roessler1992a, sears1999a}.

\textit{Permeability.} The degree of permeability of the medium in which gases diffuse will influence these internal sublimation/recondensation processes, but also the ejection in the coma of gases and dust particles. Experiments on gas diffusion have been performed by \citet{Gundlachetal2011a} to study the permeability of dust layers. Two approaches were used. The first approach utilized a gas flow streaming through a sample placed inside a glass tube. By measuring the gas flow rate and the pressure difference between the two pressure reservoirs above and below the sample, the permeability was derived. The second approach used a sublimating ice surface that was covered by a dust layer. Both experiments yielded the same results, namely that the escaping rate of molecules is reduced by the dust layers. The thicker the layer the stronger the reduction. In addition, \citet{Schweighartetal2021} studied the gas diffusion through granular materials in great detail. They varied grain size as well as the material type. The main result is that the density as well as the grain size distribution of the sample play a major role for the permeability of the sample, larger grains allowing for larger permeability values. Additional measurements in low and hyper-gravity conditions have recently been collected by \citet{capelo2022}. They show that for samples with irregularly-shaped particles the deposition and compaction of the dust layer in conditions of variable gravity have a strong influence on the measured gas permeability.   

\subsubsubsection{2.3.5.2. Chemistry involving the gas and the dust}

When the gaseous molecules produced by the sublimation of ices flow through the cometary nucleus before expanding in the coma, collisions between them and/or with the surface of ice or dust grains may lead to their adsorption followed by chemical reactions, changing the composition of the gas, and possibly forming new molecules in the ice or dust. 

\textit{Isotopic fractionation.} The influence of the sublimation process on the partition of isotopes between the solid and gas phases was studied in several laboratory experiments. During some of the KOSI cometary simulation experiments, isotopic analyses of the ice remaining at different depths after insolation under vacuum of mixtures of water, $\mathrm{CO_2}$ ices and silicate dust were performed \citep{roessler1992a, roessler1992b}. The upper layers of water ice were found to be enriched in heavy isotopes of hydrogen (deuterium, $\mathrm{D}$) and oxygen ($\mathrm{^1{}^8O}$), whereas the lower layers kept their initial isotopic composition. \citet{moores2012a} measured the isotopic composition of the water vapor produced continuously by the sublimation of water ice mixed with 1 to 25 wt\% of various types of dust grains larger than 1 $\mu$m. The $\mathrm{D/H}$ of water vapor was observed to decrease with time, particularly for samples of highest dust contents, with isotopic fractionation factor ($\mathrm{(D/H)_i{}_c{}_e / (D/H)_g{}_a{}_s}$) up to 2.5. The preferential adsorption of $\mathrm{HDO}$ on dust grains is the proposed explanation for the depletion of deuterium in the gas phase. By contrast, experiments studying the sublimation of pure water ice have reported much lower fractionation factors, varying from 0.969 to 1.123 \citep{l2017a, mortimer2018a}. The $\mathrm{D/H}$ of $\mathrm{H_2O}$ in comae has been shown to vary between 1 and 3 times the Earth’s ocean value. Based on observational data, \citet{lis2019a} suggested that $\mathrm{D/H}$ of $\mathrm{H_2O}$ in comae may be correlated with the active fraction of the comet nuclei: the more active, the lower the $\mathrm{D/H}$. \citet{lis2019a} proposed that hyperactive comets eject icy grains from deeper in the nucleus than less active comets such as 67P, outgassing higher $\mathrm{D/H}$ water from shallower depths. Future experiments on isotopic fractionation of gases emitted by comets are needed to test this scenario and explain these observations, which are of fundamental importance to constrain the origin of Earth’s water \citep{o2018a}. Experiments should be conducted with aggregates of sub-micrometer-sized dust grains, and with relatively large dust-to-ice ratio simulating the upper layer of cometary nuclei. Moreover, they should investigate not only the $\mathrm{D/H}$ fractionation of water, but also $\mathrm{^1{}^8O/{}^1{}^6O}$, $\mathrm{^1{}^7O/{}^1{}^6O}$, $\mathrm{^1{}^3C/{}^1{}^2C}$ and $\mathrm{^1{}^4N/{}^1{}^5N}$ fractionations of $\mathrm{H_2O}$, $\mathrm{CO_2}$, CO and $\mathrm{NH_3}$ molecules for example.

\textit{Alteration of mineral and organic dust.} The gas constantly diffusing through the cometary dust may also react with the dust and change its composition. \citet{takigawa2019a} exposed amorphous silicates to $\mathrm{D_2O}$ and $\mathrm{D_2O}$ + 0.15\% $\mathrm{NH_3}$ ices and vapors for 10 to 120 days, at temperatures from 246 to 323~K. Hydration of the amorphous silicates was only observed in experiments above 298~K. Experiments performed at low temperatures did not show any modification of the silicates, or of organic molecules which were exposed to the same $\mathrm{D_2O}$ and $\mathrm{NH_3}$ vapors, suggesting that alteration of dust by vapors should not be significant on comets.

\subsubsubsection{2.3.5.3. Experiments on ejections of dust/ice particles, and nucleus erosion}

Large-scale insolation experiments have reported the ejection of dust/ice particles from cometary analogue samples irradiated under vacuum \citep{thiel1989a, poch2016b, Bischoffetal2019, haack_sublimation_2021}. In these experiments, as the dust mantle left by the sublimation of the surface ices becomes thicker, it imposes more resistance to the gas flow. As the tortuosity of the path followed by the gas to exit the mantle layer increases, the force exerted by the gas on the mantle increases up to the point when it reaches the tensile strength of the mantle and breaks it in fragments that are ejected away by the gas flow (see Figure \ref{fig:haack21subli}c). The sudden release of pressure within the ice results in an increase of its sublimation rate and a decrease of its temperature because of latent heat absorption. After the ejection event, the mantle is progressively rebuilt on the surface of the ice.

In recent years, the group at the University of Warsaw has performed an extensive series of experiments to better understand how the sublimation of water ice is influenced by various parameters such as the presence, nature, concentration and location of contaminants including silicate dust and organics of variable volatility \citep{kossacki2014, kossacki2017, kossacki2019, kossacki2019b, kossacki2021, kossacki2022}. The experiments are conducted in a vacuum chamber equipped with a cooled plate where the temperature gradients within the sample (5.5 $\times$ 8 cm) and the rate of recession of the surface are measured continuously. The results of these experiments are used to constrain and validate a thermo-physical model of the surface and subsurface evolution, which can then be used to predict the evolution of the nucleus topography \citep{kossacki2018, kossacki2019c, kossacki2020b}. The case of the sliding of the silicate dust at the surface of the sublimating ice was also investigated. Experiments showed that the reduction in friction caused by the drag force from the gas results in sliding at angles much lower than the angle of repose in static conditions \citep{kossacki2020, kossacki2022}.

Other recent experiments on outgassing and material ejection have been performed by the CoPhyLab team at the Technical University of Braunschweig. Pure granular water ice samples were illuminated by an artificial Sun (\textit{Molinksi et al.}, in preparation, 2022). The energy input into the sample material caused the ejection of water ice particles from the water ice sample. The first main result of this investigation is that water ice can eject itself, which means that the pressure increase inside the ice sample occurs before the sublimation front is able to erode the overlying particle layers. In addition, the particle trajectories have been analysed and it was concluded that the ejected material possesses a non-zero starting velocity, which implies that the particles either start from within the porous sample, or have experienced a rapid initial acceleration by the ejection mechanism. The larger the temperature of the ice sample, the larger the initial velocity of the grains.

In a previous work, \citet{Bischoffetal2019} studied the ejection of dust pebbles by sublimation from an icy surface. A pure, solid ice surface was produced and dust aggregates (composed of silica) were placed on top of the ice surface. Then, the temperature was sequentially increased by a heater installed inside the ice. At temperatures of about $200 \, \mathrm{K}$ the material was ejected from the ice surface by the pressure build up at the ice-dust interface. The thicker the aggregate layer, the larger the ejected aggregate clusters are. A tentative explanation of this trend is that as the layer thickness increases, the force exerted by the gas on the dust layer increases, causing some re-arrangements of dust particles that increase the layer internal cohesion, resulting in the ejection of larger chunks.

\citet{spadaccia2022} have investigated experimentally whether the sublimation of the ice contained in mm-sized pebbles resulted in the ejection of silicate dust grains. They found that even under the harsh conditions of the laboratory (Earth gravity, accelerated sublimation), most of the initially icy pebbles retained their integrity upon sublimation of the ice while pebbles made of coarser silicate particles disrupted.

Experiments measuring the mechanical properties of various cometary surfaces analogues, described in more details in section 2.4.3, also provide knowledge on the ejection of particles from comets. The tensile strength (i.e., the pressure required to separate one particle from another) is orders of magnitudes lower for surfaces made of dust/ice pebbles (Pa) \citep{Blumetal2014} compared to homogeneous dust/ice powders (kPa) \citep{Gundlachetal2018, Bischoffetal2020}. If we assume that the gas pressure of volatile compounds are typically less than a Pascal, then ejection of materials by sublimation seems to be only possible if the surface consists of pebbles.

Finally, the comet size may influence its activity and erosion rate. \citet{Blumetal2014} have studied the compaction of pebble assemblies under simulated load of the material. A gas stream was used to compress samples positioned inside a glass-vacuum tube. In a next step, the gas stream was inverted to destroy the pebble assemblage in order to measure the low tensile strength of the packing. The main result of these experiments is that the pebbles do “\textit{remember}” $\sim 3 \%$ of the applied compression after load reduction (“\textit{memory effect}”). This effect has a major implication for understanding cometary activity. Excavated aggregate layers which have experienced gravitational compression require an enhanced gas pressure to be released. Hence, large enough comets should not be able to sustain dust activity until the whole nucleus has been eroded. \citet{Gundlachetal2016} argued that the activity stops after a threshold tensile strength is reached, because the gas pressure build up will then not be able to overcome this critical value. The interiors of smaller comets have not experienced significant gravitational compression so that the memory effect can be neglected and, in this case, the whole object can in principle be eroded by the gas pressure build up (if enough energy is supplied). Asteroids in cometary orbits (ACOs) are non-active bodies on cometary-like orbits and \citet{Gundlachetal2016} state that these objects are extinguished cometary nuclei, which were born larger, which were eroded by sublimation until the threshold value was reached, and are now inactive.

Despite recent progress on these questions, the exact mechanisms of ejection and the initial velocities of the ejected particles deserve additional investigations with a wide range of experimental parameters. The role of super- and hyper-volatiles should also be studied, starting with carbon dioxide mixed with water and dust. Such experiments have just been initiated within the CoPhyLab project.

%---------------------------------
\subsection{Experiments to constrain the composition and structure of cometary nuclei from observations}
\label{sec:subsection4}

Laboratory experimentation is crucial for the interpretation of observational data. Laboratory investigations serve two distinct and complementary purposes: (1) Laboratory measurements provide values of the fundamental physical quantities of relevant materials in their pure forms, (2) Laboratory setups provide the opportunity to prepare complex but well-controlled samples to mitigate the lack of “\textit{ground truth}”. 

\subsubsection{Surface light scattering properties: Spectroscopy, Photometry}

Over the reflected solar spectrum, cometary nuclei display an overall low reflectance, a red slope in the visible and a strong absorption around 3 $\mu$m (see the chapter by \textit{Filacchione et al.} in this volume). A number of laboratory experiments on complex analogues have attempted to reproduce these properties to decipher the chemical nature and physical properties of the darkening and red component(s) as well as the ices with which they are mixed. This empirical approach is complementary to modelling and simulation, where the measured optical constants of the pure materials are used as inputs to calculate the spectrophotometric properties of parameterized surfaces.

The optical constants of the two types of ice observed on cometary surfaces, water and carbon dioxide, have been measured at the temperatures relevant for cometary nuclei on thin films of ice deposited on cryostat windows \citep{warren1986, hansen1997, grundy1998, schmitt1998}. These values can be used to model the reflectance of macroscopic icy surfaces using a parametric representation of their properties \citep{hapke2005, shkuratov1999}, which permits the quantitative inversion of remote-sensing observations \citep{filacchione2019}.

%--------------------------------- Figure SPIPA-B
\begin{figure}[ht]
\includegraphics[width=\columnwidth]{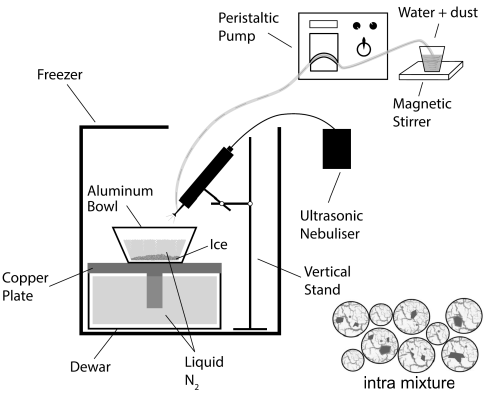}
\caption{The Setup for the Preparation of Icy Planetary Analogues (SPIPA-B) is used to produce comet analogue mixtures of ice and dust (Fig. \ref{fig:nh4residue}) \citep{pommerol2019}.}
\label{fig:spipab}
\end{figure}
%---------------------------------

Efforts have been made to produce well-characterized and reproducible particles of water ice \citep{gundlach2011, jost2013} and mix them with contaminants in different ways \citep{yoldi2015, poch2016a, pommerol2019, ciarniello2021} (Figure \ref{fig:spipab}). The same work should now be done for CO$_2$ ice and later other volatiles. While carbon and carbon-based compounds have often been used to lower the albedo of cometary analogues \citep{oehler1991, stephens1991, moroz1998}, opaque minerals such as nanoparticle Fe-Ni alloys or iron sulfide are also relevant \citep{quirico2016, rousseau2018, sultana2023}. Experimenting with complex organics and nanoparticles of metal is very challenging as the materials are difficult to procure in large amount, might be unstable in ambient laboratory conditions and/or hazardous for the experimenters. These difficulties currently limit the experimental investigations on the exact nature and properties of cometary material.

Beside the exact nature of the components, the way they are mixed with bright ices, and in which quantity, are also of primary interest. Intimate mixtures of transparent ices and fine-grained dark material show that the absorbing material has an influence disproportionate to its abundance, and minute amounts are sufficient to lower the albedo enough to hide the ice \citep{yoldi2015, jost2017a}. Unfortunately, this makes the inversion of albedo into dust-to-ice ratio highly challenging without any \textit{a priori} knowledge of the physical properties of the compounds.

In addition to the overall albedo, the spectral red slope and the dependence of the reflectance to phase angle have been investigated, as they provide additional constraints on the nature and properties of the surface material. The first attempts at reproducing the observed properties of 67P nucleus with mixtures of ice, organics and minerals have shown that while it is possible to find mixtures of ice, minerals and organics that mimic each individual property, satisfying combinations of constraints is more challenging and no analogue could be produced which fits all observed properties \citep{rousseau2018, jost2017a}. 

Sublimation experiments with icy analogues produce desiccated textures which probably resemble closely the surface of cometary nuclei. Depending on the initial composition of the dust and how it is mixed with the ice, sublimation can produce high-porosity cohesive mantles which display unique spectro-photometric properties \citep{poch2016a, poch2016b}. The strength of absorption bands is reduced by the decrease of size and the deaggregation of the dust grains, to a point where spectral features are barely recognisable despite a high abundance of the material \citep{sultana2021, sultana2023}. The strong backscattering behaviour of such porous mantle is the best match to the phase dependence of reflectance observed at the nucleus of 67P \citep{jost2017b}. In addition to the micro-porosity, the sub-micrometer grain size of components having contrasted optical indexes (such as mixtures of silicates and opaque minerals or organics) could explain together the low albedo, visible and near infrared spectral slope, mid-infrared spectral emissivity, and visible polarimetric phase curves of the most primitive small bodies, as shown in \citet{sultana2023}. While laboratory measurements are useful to understand the photometry of the nucleus at low phase angle, large-scale roughness and topographic effects likely dominate the photometry at high phase angle, which cannot easily be studied in the laboratory. Models and/or very large-scale experiments would be required to make progress on this question.

\begin{figure*}
\begin{center}
\includegraphics[width=17.1cm]{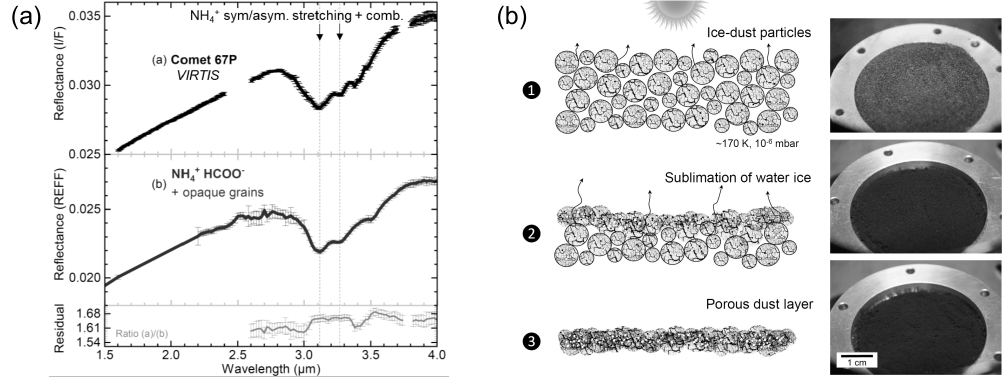}
\caption{ Identification of $\mathrm{NH_4^+}$ salts on the surface of comet 67P. (a) Comparison of Rosetta/VIRTIS average 67P spectrum with the one of a sublimation residue produced by (b) mixing water, $\mathrm{NH_4^+ HCOO^-}$ and opaque sub-µm grains to make icy-dust particles (Fig. \ref{fig:spipab}), followed by the sublimation of the water in a vacuum chamber \citep{poch2020}.}
\label{fig:nh4residue}
\end{center}
\end{figure*}

Beside the visible red slope, the strong 3-$\mu$m band of 67P is the most prominent spectral feature identified on a cometary nucleus. Its interpretation is challenging as different components contribute to it. Here as well, sublimation experiments have proven crucial to measure spectra influenced by both the chemical nature of the compounds and the physical texture of the surface. \citet{poch2020} were able to prove in this way for the first time that ammonium salts are a key component of the surface of 67P (Figure \ref{fig:nh4residue}), while \citet{raponi2020} have showed the additional contribution to the 3-$\mu$m band of aliphatic organics, and \citet{mennella2020} the possible contribution of hydroxylated silicates.

%--------------------------------- Figure Poch2020 - NH4residue
%---------------------------------

To be useful to the large community, the laboratory data (optical constants, reflectance spectra, samples descriptions etc.) need to be available readily and freely. Recent efforts have resulted in the development, update and modernization of numerical databases to store and distribute these data and associated documentation \citep{schmitt2018, milliken2016, kokaly2017}.  

During approach to the Sun, cometary surfaces are irradiated by solar wind and UV light, and interactions of coma gases with the solar wind generate accelerated molecular ions which impact and sputter the nucleus surface, as seen on comet 67P \citep{nilsson2015, wurz2015}. The influence of energetic particles on the optical properties of cometary surfaces are studied by experiments described in section 2.2.2. The irradiation dose received during approach to the Sun may be lower than during their residence in the outer solar system, because of the higher resurfacing rate due to cometary activity, and lower energy of SEPs compared to GCRs. Finally, we note that the charging of the cometary surface may induce the levitation of dust particles as shown in the laboratory by \citet{wang2016}, possibly influencing surface optical properties.

\subsubsection{Subsurface and interior electrical properties}

The quantitative interpretation of radar data relies on the knowledge of the complex dielectric constants of the materials encountered at the surface and in the subsurface of cometary nuclei. While the dielectric constants of many materials of interest in their pure form have been determined for decades (see Table A1 in \citet{herique2016}), recent experiments have focused on the measurements of dielectric properties of complex synthetic or natural analogues for cometary material. Indeed, as for other spectral domains, mixing laws exist and are readily used \citep{sihvola2000} but their applicability to complex high-porosity mixtures of ices, minerals and organics must be verified. 

Motivated by the interpretation of multi-instrumental Rosetta and Philae measurements at comet 67P, \citet{brouet2016a} have measured systematically and over a wide spectral range the dielectric permittivity of porous mixtures of water ice and dust. The application of these measurements to Rosetta data indicates a significant vertical gradient of porosity in the 67P nucleus \citep{brouet2016b}. 
To interpret the data from the CONSERT instrument, \citet{heggy2012} and \citet{Kofman2015} reported on new measurements of the dielectric properties of some carbonaceous chondrites, thought to be reasonable analogues for the dust found in cometary nuclei. The values are notably lower than values for ordinary chondrites found in databases, but the authors also note the need to better study experimentally the potential effect of metal oxidation which might bias the results \citep{herique2016}.
Beside Titan, the dielectric measurements of tholins performed by \citet{lethuillier2018} are also relevant for cometary material in which relatively heavy and complex organic molecules are abundant. The measurements revealed that both the real and imaginary parts of their dielectric constant are sensitive to the composition of the gas from which tholins are produced.

\subsubsection{Experiments on mechanical properties of dust/ice surfaces}

Laboratory studies have been performed to measure the tensile strength of granular materials. Two different material types have to be distinguished. The first type comprises all samples whose structure can be described as a homogeneous dust layer. These experiments can be used to understand the tensile strength of dust layers that were formed by the mass transfer scenario (see section 2.1.1 and the chapter by \textit{Simon et al.} in this volume), and can be used to investigate the internal tensile strength of the dust/ice pebbles. \citet{Gundlachetal2018} have performed experiments (so-called “\textit{Brazilian Disc Tests}”) to study the tensile strength of cylindrical samples a few centimeters large, composed of granular silica, or water ice. A loading platen is lowered on to a cylindrical sample by a stepper motor and exerts a force and, thus, a pressure on to the cylinder. The sample stays intact until the tensile strength of the material is reached. These measurements have shown that the tensile strength of homogeneous samples is in the $\mathrm{kPa}$ range and no difference between water ice and silica was detected. This implies that the internal tensile strength of pebbles as well as the tensile strength of cometary surface formed by the mass transfer process are in the $\mathrm{kPa}$ range. Also, different organic materials have been studied with the result that the exact tensile strength values deviate from sample to sample, but all investigations yield values of about $10^{-1} \,\mathrm{kPa}$ to $10^{1} \,\mathrm{kPa}$ \citep{Bischoffetal2020}. In contrast, \citet{Blumetal2014} studied the tensile strength of surfaces made of pebbles made of ice and dust, and found that their tensile strength is orders of magnitude lower, namely in the Pa range. These experiments are crucial for our understanding of cometary activity discussed in section 2.3.5.3, because if we assume that the gas pressure of volatile compounds is typically less than a Pascal, then ejection of materials by sublimation seems to be only possible if the surface consists of pebbles. In other words, comets that show activity (ejection of dust) must possess a pebble-like surface structure with a very-low tensile strength.

Following many large-scale insolation experiments (described in section 2.3.2), a hardening of the remaining ice was observed underneath the dehydrated dust mantle at the surface \citep{kochan1989, ratke1989, seiferlin1995, thiel1989a, poch2016b, kaufmann2018} (Figure \ref{fig:rickman}a). This consolidation can be explained by the re-condensation of $\mathrm{H_2O}$/$\mathrm{CO_2}$ between the individual ice/dust particles and/or by sintering: both processes build bonds connecting the ice particles together and providing solidification of the sample. Measurements of the hardness of these ice-dust crusts, performed after the KOSI experiments, yielded values of 0.15 to 5 MPa \citep{gruen1993a, kochan1989}. Experiments also showed that the hardening depends on the amount of dust \citep{kaufmann2018}, and on the way the ice and the dust are initially mixed together \citep{poch2016b}.

While a number of experiments with cometary analogues have provided observations on the formation and evolution of morphologic features upon sublimation of the ice, few laboratory simulations have focused entirely on this aspect. \citet{haack_sublimation_2021} and \citet{haack2021b} have recently investigated the evolution of mixtures of water ice and dust (fly ash) under simulated cometary conditions and using scaling laws to select adequate particle sizes and energy input taking into account the effect of terrestrial gravity. Figure \ref{fig:haack21subli}c shows an example of a sample affected by the ejection of large agglomerates. The morphologies observed are diverse, some of them reminiscent of the high-resolution images obtained at 67P such as fractures, overhangs and collapses. Both the dust-to-ice mass ratios and the direction of the incident light strongly affect the morphology resulting from the sublimation. The introduction of salt (sodium acetate) and of an amino acid (glycine) in the sample \citep{haack2021b} changes the outcome of the experiments by inhibiting the collapse events, leading to thick, cohesive and stable but fractured mantles at the surface of the sample.

The influence of composition on the sublimation coefficient, and hence the erosion rate of cometary analogues, has been studied experimentally by the Warsaw group (see \citet{kossacki2014} and other references listed in section 2.3.5.3). Based on these experimental results, they have developed a thermal model, which they then use to interpret the topography observed on cometary nuclei \citep{kossacki2018, kossacki2019c, kossacki2020b}.

%-----------------------------------------------------------------------------------------------------------
\section{\textbf{COMAE}}
\label{sec:anothersection1}

\subsection{Experiments to constrain the composition and structure of comae}
\label{sec:subsection6}

Comae are made of gases dragging solid particles of dust and ice. They are observed remotely via different techniques. Ultraviolet-visible, infrared or radio/mm-wave spectroscopy allow to determine the composition of the gas phase. Broad-band visible photometry, polarimetry, and infrared spectroscopy are used to infer the composition and physical properties of the solid particles.

\subsubsection{Coma gases}
 Gases produced by the sublimation of ices inside the nucleus may undergo several processes modifying their composition before they reach the coma, as discussed in section 2.3.5.2. Once in the coma, these molecules collide with each other, with electrons and solar wind ions. All these processes and their resulting spectral emission need to be studied in the laboratory and computationally to understand fully and correctly the light emitted by comae. This requires the determination of physical parameters such as frequencies and strengths of lines, collisional cross-sections, photodissociation rates etc. for many molecules and their isotopologues, which are reported in databases. These laboratory experiments and databases on gas coma properties are reported in the chapters by \textit{Bodewits et al.} and \textit{Biver et al.} in this volume. In the following paragraphs, we will focus on experiments on solid particles in the coma.

%--------------------------------- Figure Lightscat_progra2
\begin{figure*}
\begin{center}
\includegraphics[width=17.1cm]{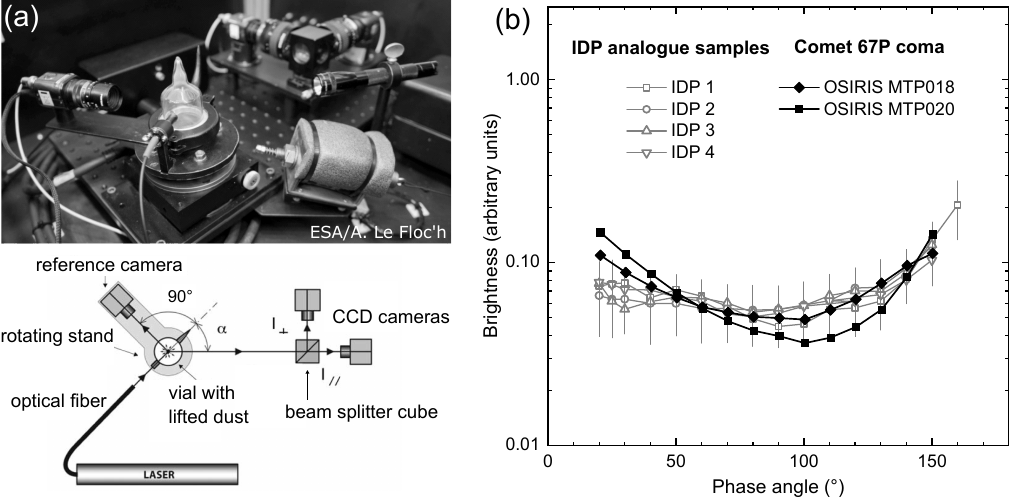}
\caption{Light scattering properties of cometary dust analogues in suspension. (a) PROGRA2-Vis instrument (top) picture of the setup under microgravity, (bottom) scheme of laboratory setup \citep{hadamcik2009b}. (b) Comparison between phase functions of comet 67P (Rosetta/OSIRIS data) and measurements of IDP analogues (aggregates of sub-$\mu$m grains of carbonaceous compounds and minerals, such as those shown in Fig. \ref{fig:organicgas} and \ref{fig:smoke}) \citep{levasseur-regourd2019}.}
\label{fig:lightscat}
\end{center}
\end{figure*}
%---------------------------------

\subsubsection{Coma particles' light scattering properties}

Dust in the coma can be studied from a distance by analysing thermally emitted light at long infrared wavelengths and solar scattered light in the near-UV to near-IR spectral range (see the chapter by \textit{Kolokolova et al.} in this volume). In the first case, the spectral energy distribution (SED) of the emitted light is the primary source of information on the grains properties. At shorter wavelengths, the analysis of the angular distribution of scattered light can provide additional information on the physical characteristics of the dust beside the spectral dimension. The study of the light polarization ideally complements total light analyses for that purpose.

The interpretation of infrared SEDs relies on laboratory measurements of the thermal emissivity of fine powdered materials relevant for comets, such as minerals, carbonaceous compounds and ices, essential to infer the composition, temperature, size distribution and porosity of cometary dusts \citep{lisse2007a, wooden2008a}. The numerous challenges related to the selection and procurement of suitable analogue materials discussed in the context of the nucleus are also relevant here.

Analyses of the scattering of light by dust particles in the coma can provide quantitative information on a number of properties of the dust, but the inversion of observational data into physical characterization of the particles is a complicated process which relies on scattering models and parameterizations of the particles which should be tested (see the chapter by \textit{Kolokolova et al.} in this volume). As unpolarised Sunlight gets partially polarized when scattered by particles in the coma, it carries information on the complex refractive index, i.e. the composition, as well as on the size, shape and structure of the dust particles, complementary to SED observations \citep{kolokolova2007a}. The variations of polarization with wavelength and phase angle, as comets orbit the Sun, provide information on these properties \citep{kiselev2015a, hines2016a} and are therefore extensively studied in the laboratory to compare with the observations \citep{levasseur-regourd2015a}. Light scattering properties of particles randomly oriented in a gas stream are measured at the Cosmic Dust Laboratory (CoDuLab) of the Instituto de Astrofisica de Andalucia (IAA) in Granada \citep{hovenier2009a, munoz2010a, munoz2011a}, where the whole Mueller scattering matrixes \citep{munoz2012a} have been obtained for a variety of analogues. Phase functions of millimeter-sized irregular dust grains have been compared to numerical models with the aim to improve the interpretation of comets and protoplanetary disk observations \citep{munoz2017a, escobar-cerezo2017a}. The PROGRA2 instruments have been developed in France since 1998, in order to compare linear polarization phase functions of small bodies surfaces and cometary comae with those measured on a wide set of samples \citep{levasseur1998, worms1999}. Measurements of interest for comae may be obtained in the laboratory for particles smaller than 10 µm through an air-draught technique, while more systematic measurements are obtained under microgravity conditions (0 $\pm$ 0.05 g) during parabolic flight campaigns organized by CNES or ESA space agencies \citep{hadamcik2011} (Figure \ref{fig:lightscat}). The PROGRA2-Vis instrument can provide measurements between 5° ad 165° phase angles at 543.5 nm and 632.8 nm, while the PROGRA2-IR instrument operates in the IR, at 1.5 $\mu$m \citep{levasseur-regourd2015a}. The polarimetric phase curves of cometary analogues made of porous aggregates of sub-micrometer-sized Mg-silicates, Fe-silicates and carbon black grains mixed with compact Mg-silicates grains appear similar to observations of comae \citep{hadamcik2007a}. Aggregates of sub-micrometer-sized monomers provide the best fits for higher polarization observed in cometary jets, while a mixture of porous aggregates and compact grains is needed to fit whole comae observations \citep{hadamcik2006a}. The influence of the size of individual dust grains (from nm to $\mu$m) and agglomerates (from $\mu$m to cm) on the maximum polarization of phase curves has been studied \citep{hadamcik2009b}, allowing the interpretation of polarimetric images of the coma of 103P/Hartley~2 in term of progressive sublimation of ice, leading to particle fragmentations with increasing distance from the nucleus \citep{hadamcik2013a}. Finally, a light scattering unit, derived from PROGRA2-Vis, had been developed as a tentative precursor experiment for aggregation of microspheres in microgravity conditions, and flown on the ESA sounding rocket flight MASER-8 \citep{levasseur2003}.

\subsection{Experiments on the processes transforming coma particles}
\label{sec:subsection7}

The dust particles ejected in the coma are subjected to different processes (photolysis, heating, etc.), which can modify their composition and generate additional gaseous products. As a consequence, the distribution of some gaseous species in comae correspond not only to their direct sublimation from the nucleus, but also to their production from larger molecules or dust grains. Such additional sources of gas in the coma are called “\textit{extended sources}” or “\textit{distributed sources}” \citep{cottin2008a}. Distributed sources are observed for molecules and radicals such as $\mathrm{H_2CO}$, $\mathrm{CO}$, $\mathrm{HNC}$, $\mathrm{CN}$, $\mathrm{OCS}$, $\mathrm{HF}$, $\mathrm{HCl}$, $\mathrm{C_2}$, $\mathrm{C_3}$ and even glycine (see the chapter by \textit{Biver et al.} in this volume).

The absorption of solar radiation by molecules in the dust can cause their photo-degradation by UV photons, and it also induces the warming of the dust. Volatile and semi-volatile species still present in the particles can undergo photolysis or sublimation, and even refractory compounds can decompose into gases due to irradiation or heating. Indeed, the sub-micrometer-sized particles made of absorptive materials can reach temperatures higher than a black-body, up to 500 K or more for 0.2 $\mu$m grains \citep{kolokolova2004}. In addition to solar radiation, dust particles could also be irradiated by energetic particles from the cometary plasma and solar wind \citep{mendis2013a}.

%--------------------------------- Figure Fray06
\begin{figure}[ht]
\includegraphics[width=\columnwidth]{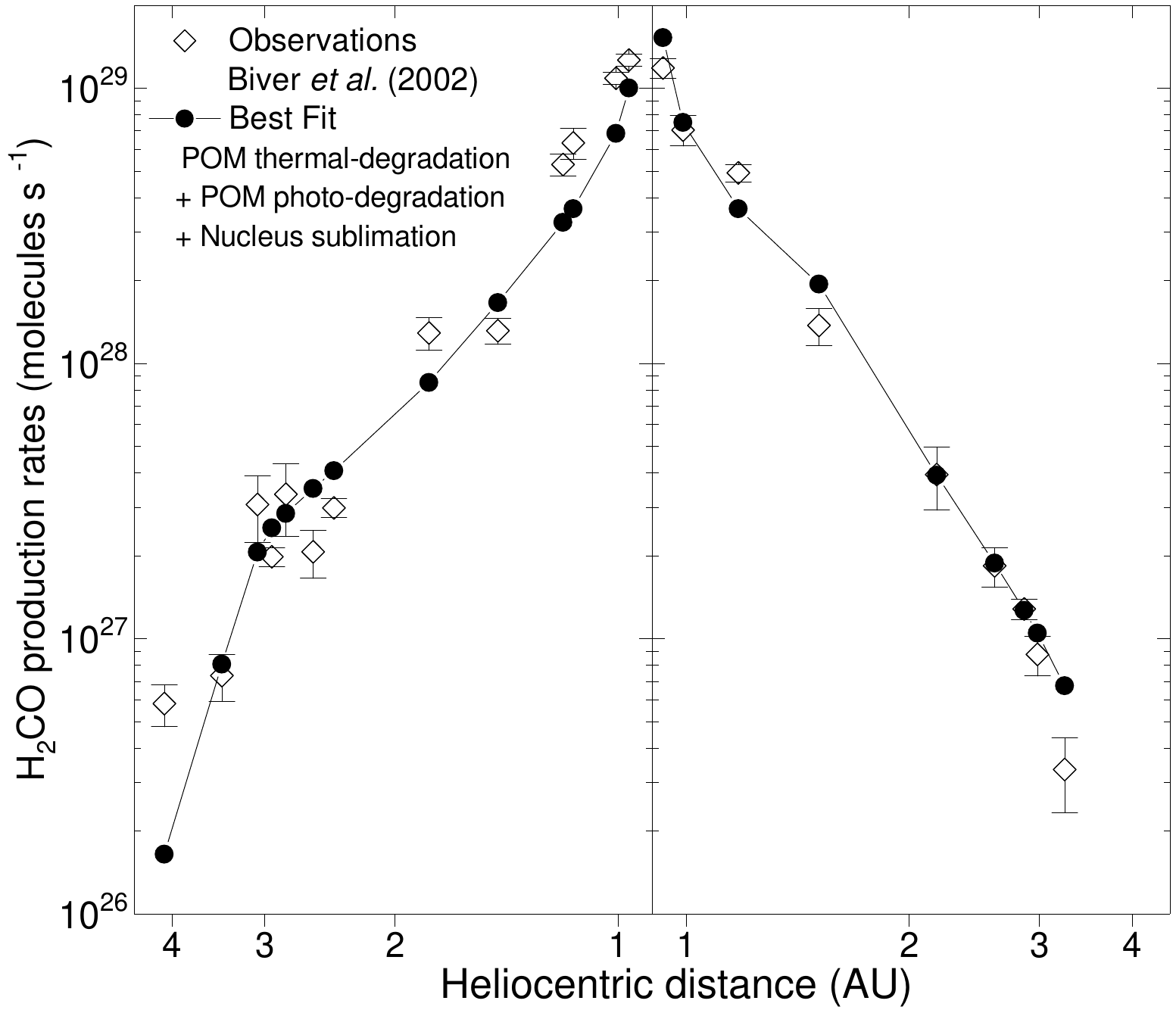}
\caption{$\mathrm{H_2CO}$ production rate as a function of heliocentric distance for comet C/1995 O1 Hale-Bopp as observed (open squares), and as computed from experimental data on POM (black circles), assuming a mass fraction in the grains of 3.1\% and $\mathrm{H_2CO}$ production at the surface of the nucleus equal to 3\% of $\mathrm{HCN}$ production \citep{fray2006}.}
\label{fig:fray06}
\end{figure}
%---------------------------------

Laboratory experiments are performed to identify the nature of the solid precursors responsible for these distributed sources, and retrieve qualitative and quantitative data on the processes by which they are transformed in gases. Experiments investigating photo-degradation and thermal-degradation of compounds relevant for comets enable the determination of absorption cross sections, quantum yields of photo-degradation, Arrhenius constants and activation energies of thermal-degradation, or vapor pressures of sublimation, which serve as inputs into numerical models of gas distributions observed in comae \citep{cottin2008a, hadraoui2019a}. Photolysis of solid compounds stable at room temperature can be performed in a temperature-controlled (300~K) reactor illuminated with a UV lamp, maintained under vacuum ($\mathrm{10^{-^4}}$ mbar) and connected to a system analyzing the gaseous products, such as an infrared spectrometer or a mass spectrometer possibly coupled with a gas chromatograph \citep{cottin2008a, fray2004b}. Thermal degradation of the same kind of compounds can be performed in a heated reactor or a pyrolyser connected to similar analytic systems. To study the degradation of compounds formed after irradiation and heating of cometary ices, high vacuum chambers such as the ones described in section 2.1.2.1 and Figure \ref{fig:icechem} are used to produce these compounds and degrade them \textit{in situ} via UV and/or controlled heating from 20 to 800 K \citep{briani2013a}. Several experiments have studied the photo-degradation and/or thermal-degradation of refractory organic compounds such as polyoxymethylene (POM, $\mathrm{[(-CH_2O-)_n]}$) \citep{fray2004a, fray2006} (Figure \ref{fig:fray06}) and hexamethylenetetramine (HMT, $\mathrm{[(CH_2)_6N_4]}$) \citep{briani2013a, fray2004b} produced from ice chemistry (see section 2.1.2.2), HCN-polymers \citep{fray2004b}, and $\mathrm{C_3O_2}$-polymers which have been proposed as a source of $\mathrm{CO}$ (see references in \citet{cottin2008a}). However, to date the presence of these compounds in cometary dust is uncertain. The Rosetta mission has shown the presence of high-molecular-weight organic matter sharing similarities with meteoritic IOM \citep{fray2016a, raponi2020} and ammonium salts \citep{altwegg2020a, poch2020} in the dust of comet 67P. Therefore, future experiments should investigate the photo- and thermal-degradation of these types of compounds. The thermal decomposition of ammonium salts ($\mathrm{NH_4^+Cl^-}$ and $\mathrm{NH_4^+HCO_2^-}$) and the detectability of their products after ionization in the Rosetta mass spectrometer ROSINA was studied in the laboratory, enabling their identification \citep{haenni2019a}. Moreover, $\mathrm{NH_4^+}$ salts thermal degradation at relatively high temperature could explain the apparent depletion of comae in nitrogen, the increase of $\mathrm{NH_3}$ in comets closer to the Sun \citep{altwegg2020a, dello2016a}, and possibly several distributed sources \citep{haenni2020a}. To confirm this hypothesis, future experiments of photo/thermal-degradation on pure and mixed ammonium salts should be performed.

The dust in comae may be subjected to other processes not yet investigated in experiments. In particular, the solar wind interaction with the coma, responsible for the cometary ionosphere, could induce radiolysis of dust particles by energetic particles or charging processes, causing erosion and disruption of dust particles \citep{mendis2013a}.

\subsection{Link with circumstellar disks}

Our knowledge of circumstellar disks has evolved considerably in the last decades thanks to fast progress in observations at optical and radio wavelengths. These data provide new and important observational constraints to better understand planetary formation and complement knowledge gained in our solar system by studying primordial objects such as comets. Indeed, circumstellar disks are made of objects whose physico-chemical properties could be similar to cometesimals or comets depending on the disk age, and studies of circumstellar disks and comets share the same scientific questions on the origin and evolution of matter around stars. For these reasons, experiments on comets are also useful to better interpret observations of circumstellar disks (see the chapters by \textit{Aikawa et al.} and \textit{Fitzsimmons et al.} in this volume). \citet{Levasseur-Regourd2020} show that the properties of cometary dust, in particular their high porosity, are compatible with modeling of light scattering in protoplanetary and debris disks. For example, \citet{hunziker2021} extract the polarimetric properties of dust particles in the protoplanetary disc HD 142527 and conclude that large ($>$ 1 $\mu$m) and porous aggregates are present. Future laboratory efforts to further characterize light scattering properties of analogue samples relevant for circumstellar disks will be necessary to better constrain the physical models on which these interpretations are based. While the initial steps can largely be based on similar experiments related to comets, some specificity of disks will also need to be taken into account.

%-----------------------------------------------------------------------------------------------------------
\section{\textbf{PERSPECTIVES FOR FUTURE EXPERIMENTS}}
\label{sec:anothersection8}

Laboratory experiments provide essential reference data to understand what comets are made of, how they formed, how they evolve, and what causes their activity. Experiments extend to all the cometary materials (ice, dust, gas) and all the processes they undergo from their formation/processing in pre-cometary environments to their incorporation, and possible transformation in comet nuclei and comae. In this chapter, we have provided an overview of the diversity of laboratory experiments supporting the scientific investigations of comets.

Over the last two decades, cometary missions (Stardust sample-return, Deep Impact, Rosetta) and astronomical observations have provided new constraints on comet structure and composition. These data should drive the preparation of future experiments, indicating specific structures, compounds and conditions of interest to study with “\textit{simple}” or “\textit{complex}” samples. In particular, experiments performed on “\textit{complex cometary analogues}” need to be continued with samples more representative of the current knowledge on both the physical structure and the chemical composition of cometary materials. For example, experiments with mixtures of ices and dust are essential to study the chemistry (formation, transformation) and the spectroscopy of cometary materials. The dust comprises not only silicates and other minerals, but also a large fraction of high-molecular-weight organic matter, as well as semi-volatile components such as lower molecular weight organic molecules and salts. These components appear to be arranged in the form of porous or compact aggregates of sub-micrometer-sized grains. Cometary analogues consisting of such realistic mixtures, in terms of structure and composition, will provide a better understanding of various physical processes such as coagulation, accretion, thermal evolution, outgassing, light scattering etc. Furthermore, the realization of these experiments in the most realistic conditions (low temperatures, vacuum, microgravity) will also be beneficial. Closer collaborations between cometary physicists and chemists should be encouraged to perform such experiments. 

Large-scale cometary laboratory experiments can also be used to explore and test new ideas of cometary processes still unidentified, such as the effects of cosmic ray irradiation in thick sections of cometary analogues of various porosity, the processes related to the diffusion of gas through the nucleus material (transformations, isotopic fractionation), the role of super- and hyper-volatiles in the ejection of dust/ice particles, or the photo/thermal decomposition and interaction with plasma of grains in comae.

Future experiments, either refining already known cometary processes or revealing new ones, will be complementary to theoretical models simulating these processes. Many open questions about comets would only receive reasonable answers via the interplay of models and experiments. For example, to understand comet formation processes it is essential to have good scaling laws to extrapolate results from experiments to large objects. Moreover, models of (pre-)cometary ice chemistry using networks of reactions (whose parameters are determined by theories and/or experiments) could allow to answer how, when and where some cometary materials were formed. Furthermore, models of comet thermal evolution and outgassing would benefit from measurements of the thermal conductivity of refractory organics, salts, and gas-laden amorphous ice, as well as measurements of the solid-gas equilibrium of ices and clathrate hydrates at low pressure and temperature. Finally, future experiments could also benefit from new technologies or methods of physical or chemical measurements, providing better characterization of the samples and of their processing.

The next cometary space missions (see the chapter by \textit{Snodgrass et al.} in this volume) will also drive the realization of future experiments. Firstly, experiments are needed to prepare and interpret these observations. In particular, one of the challenges of observing a dynamically-new comet, as planned by the Comet Interceptor mission, will be to disentangle the origins of the observed compounds or structures (already formed in cometesimals, or produced later). Secondly and reciprocally, new observations will provide knowledge to take into account in future experiments. A long term challenge of cometary science is to plan and execute future analyses of cometary nucleus material, either \textit{in situ} or via sample return. A potential future sample return of cometary nucleus to Earth would require various experimental and theoretical simulations to define how to best sample, preserve, and analyze the cometary material. Future large-scale experiments could play a fundamental role in refining our knowledge of the depth profile of a cometary nucleus (composition and interrelationship of volatiles and refractories, physical structure, thermal and mechanical properties), crucial for all steps of the mission, from sampling to interpreting the results \citep{thomas2019}. In return, such analyses of cometary material will provide excellent data to improve experimental and theoretical models of comets.

Astronomical observatories (JWST, ELT, ALMA etc.) will also provide observations of a multitude of comae and nuclei, and allow comparisons with pre- and proto-stellar environments, proto-planetary or debris disks, but also TNOs and asteroids of our own solar system. Cosmo-materials and cosmochemistry studies could provide as well interesting insights related to comets. Laboratory experiments will be instrumental in understanding the evolutionary links between these environments/objects and comets.

%--------------------------------- 
\vskip .5in
\noindent \textbf{Acknowledgments.} \\
The authors dedicate this chapter to the memory of Prof. Anny-Chantal Levasseur-Regourd (1945-2022). ACLR was a renowned specialist of comets, especially of the optical properties of cometary and interplanetary dusts she studied via observations, laboratory and numerical simulations. The first version of this chapter benefited from her thorough reading and advice. Her inexhaustible enthusiasm will continue to inspire us. The work of O.P. related to comets was supported by the Center for Space and Habitability of the University of Bern, the National Centre of Competence in Research (NCCR) PlanetS supported by the Swiss National Science Foundation, the Centre National d’Etudes Spatiales (CNES), the French Agence Nationale de la Recherche (program Classy, ANR-17-CE31-0004) and the European Research Council (under the SOLARYS grant ERC-CoG2017-771691). O.P. acknowledges Bernard Schmitt for discussions about cometary physics, and Mathieu Choukroun for discussions on the chapter's structure. The authors are grateful to Joseph A. Nuth and Perry A Gerakines for their reviews of this chapter.

%---------------------------------
\vskip .5in
\bibliographystyle{sss-three.bst}
\bibliography{refs.bib}

\end{document}